%% file: main.tex
\documentclass[twocolumn,twocolappendix,usenames,dvipsnames]{aastex63}

\usepackage{amsmath}

\shorttitle{CR Diffusion Suppression in Star-Forming Regions and Clump Formation in Gas-Rich Galaxies}
\shortauthors{Semenov, Kravtsov, Caprioli}
\graphicspath{{figures/}}

\input{citation_fix}
\input{defs.tex}

\begin{document}

\title{Cosmic-Ray Diffusion Suppression in Star-forming Regions\\ Inhibits Clump Formation in Gas-rich Galaxies}

\author[0000-0002-6648-7136]{Vadim A. Semenov}
\altaffiliation{\href{mailto:vadim.semenov@cfa.harvard.edu}{vadim.semenov@cfa.harvard.edu} \\ NHFP Hubble Fellow.}
\affiliation{Center for Astrophysics $|$ Harvard \& Smithsonian, 60 Garden St, Cambridge, MA 02138, USA}

\author[0000-0003-4307-634X]{Andrey V. Kravtsov}
\affiliation{Department of Astronomy \& Astrophysics, The University of Chicago, Chicago, IL 60637, USA}
\affiliation{Kavli Institute for Cosmological Physics, The University of Chicago, Chicago, IL 60637, USA}
\affiliation{Enrico Fermi Institute, The University of Chicago, Chicago, IL 60637, USA}

\author[0000-0003-0939-8775]{Damiano Caprioli}
\affiliation{Department of Astronomy \& Astrophysics, The University of Chicago, Chicago, IL 60637, USA}

\begin{abstract}
Observations of the $\gamma$-ray emission around star clusters, isolated supernova remnants, and pulsar wind nebulae indicate that the cosmic-ray (CR) diffusion coefficient near acceleration sites can be suppressed by a large factor compared to the Galaxy average. We explore the effects of such local suppression of CR diffusion on galaxy evolution using simulations of isolated disk galaxies with regular and high  gas fractions. Our results show that while CR propagation with constant diffusivity can make gaseous disks more stable by increasing the midplane pressure, large-scale CR pressure gradients cannot prevent local fragmentation when the disk is unstable. In contrast, when CR diffusivity is suppressed in star-forming regions, the accumulation of CRs in these regions results in strong local pressure gradients that prevent the formation of massive gaseous clumps. As a result, the distribution of dense gas and star formation changes qualitatively: a globally unstable gaseous disk does not violently fragment into massive star-forming clumps but maintains a regular grand-design spiral structure. This effect regulates star formation and disk structure and is qualitatively different from and complementary to the global role of CRs in vertical hydrostatic support of the gaseous disk and in driving galactic winds.
\end{abstract}


\section{Introduction} \label{sec:intro}

There is broad consensus that star formation and its quenching, as well as gas outflows from galaxies, are regulated by energy and momentum injection from young massive stars and feedback from supermassive black holes \citep[e.g.,][]{Somerville.Dave.2015,Naab.Ostriker.2017, Vogelsberger.etal.2020}. Details and the relative role of different feedback processes, however, are still actively debated \citep[e.g.,][]{Zhang.2018}. In particular, cosmic rays (CRs) accelerated at strong shocks formed by stellar winds and supernova (SN) explosions that accompany star formation have been the focus of much recent research. 

Indeed, CRs constitute a significant fraction of the interstellar medium (ISM) pressure budget, and therefore, they must be dynamically important. They play a key role in regulating thermal balance in dense molecular clouds and setting conditions for star formation \citep[e.g.,][]{Papadopoulos.Thi.2013}. Due to their long cooling times, CRs can significantly prolong the Sedov--Taylor stages of SN remnants, leading to larger momentum injected into the ISM \citep{diesing+18}. On larger scales, CRs can be one of the important drivers of galactic winds as suggested by a number of analytical models \citep[][see \citealt{Zweibel2017} for a  review]{Ipavich1975,Breitschwerdt.etal.1991,Breitschwerdt.etal.1993,Zirakashvili.etal.1996,Everett.etal.2008,Socrates.etal.2008,Samui.etal.2010,recchia+16}. 

Numerical simulations that included CR injection, cooling, and propagation in local ISM patches \citep{Hanasz.etal.2013,Peters.etal.2015,Girichidis.etal.2016,Simpson.etal.2016}, isolated galaxies \citep{Uhlig.etal.2012,Booth.etal.2013,Salem.Bryan.2014,Pakmor.etal.2016,Ruszkowski.etal.2017,Wiener.etal.2017,Farber.etal.2018,Holguin.etal.2019,Bustard.etal.2020,Dashyan.Dubois.2020}, and cosmological simulations of galaxy formation \citep{Wadepuhl.etal.2011,Salem.etal.2014,Chen.etal.2016,Chan.etal.2019,Buck.etal.2020,Hopkins.etal.2020c,Hopkins.etal.2020b,Hopkins.etal.2020a} support the idea that CRs may play a significant role in regulating star formation and driving winds  \citep[although CR wind driving may possibly be limited to halos of mass $M_{\rm h}\lesssim 10^{12}\, \Msun$; e.g.,][]{Fujita.MacLow.2018,Buckman.etal.2020}. CR-driven outflows likely play a significant role in shaping the properties of the circumgalactic medium around galaxies \citep[][]{Booth.etal.2013,Liang.etal.2016,Salem.etal.2016,Butsky.Quinn.2018,Girichidis.etal.2018,Ji.etal.2019} and in regulating plasma cooling in the core regions of galaxy groups and clusters \citep[e.g.,][]{Guo.Oh.2008,Su.etal.2020}. 

The key condition for CR-driven winds is the ability of CRs to propagate away from their injection sites into the inner halo of their parent galaxies \citep[e.g.,][]{Grenier.etal.2015}. By escaping from the ISM into the less dense inner halo, CRs avoid losing most of their energy to cooling and establish a significant large-scale pressure gradient in the halo that drives global wind \citep[][]{Uhlig.etal.2012,Booth.etal.2013,Salem.Bryan.2014}. 
Given the importance of CR propagation, the effects of different treatments of this process have been the focus of many recent studies. As we discuss below in Section \ref{sec:crprop}, there are a number of processes that can affect CR propagation, and theoretical understanding of these processes and their relative role remains poor. Thus, different recent studies considered the effects of different assumptions about CR propagation on the star formation and gas outflows from galaxies usually aiming to bracket the possibilities.  

For example, \citet{Pakmor.etal.2016} explored the effects of anisotropy in the CR diffusion using information about magnetic fields in their simulations of an isolated galaxy. To bracket the effect of diffusion anisotropy, they contrasted a simulation with isotropic and constant diffusion coefficient and a simulation in which diffusion (with the same constant coefficient) was allowed only along the local direction of the $B$ field. Given that diffusion was restricted in the anisotropic case, the escape of CRs from the ISM and the onset of CR-driven wind were also delayed. The larger residence time of CRs in the gaseous disk also resulted in larger cooling losses, leaving less CR energy for wind driving. 

\citet{Ruszkowski.etal.2017} showed that when CR can stream away from the ISM at velocities larger than the Alfv\'en speed, $\vA=B/\sqrt{4\pi\rho}$, the wind mass flux and mass-loading factor are enhanced. 
\citet{Wiener.etal.2017} also  considered the relative effects of CR streaming and diffusion and argued that CR propagation and wind driving can be affected by CR energy losses due to wave generation by the streaming instability \citep[see also][]{Buck.etal.2020}. At the same time, \citet[][]{Chan.etal.2019} argued that propagation that assumes only CR advection with the gas or streaming at a trans-Alfv\'enic velocity, $\vst\sim\vA$, results in a significant overestimation of the $\gamma$-ray luminosity associated with the pion production by CR interactions with thermal gas \citep[see also][]{Hopkins.etal.2020a}. With the inclusion of diffusion with a sufficiently large diffusion coefficient, observational constraints can be satisfied, but in this case, the effect of streaming with $\vst\sim\vA$ on wind driving becomes subdominant.

More recently, \citet{Hopkins.etal.2020b,Hopkins.etal.2020a} explored a wide range of models with isotropic and anisotropic diffusion and/or streaming and propagation coefficients varying with the local state of gas and magnetic fields. These authors concluded that most models can produce results consistent with the CR observations in the solar system and $\gamma$-ray measurements in other galaxies, although in many models propagation coefficients need to be adjusted by a significant factor from the values commonly assumed as fiducial. These results illustrate significant current uncertainties in the CR propagation modeling and a limited constraining power of current observations. 

In this paper, we explore a relatively simple isotropic diffusion for CR propagation, with the diffusion coefficient and CR cooling strongly suppressed near their injection sites in star-forming regions. Such suppression is motivated by several observations and theoretical arguments and modeling results, as we discuss in Section~\ref{sec:crprop}. At the same time, its effects on galaxy evolution have not been explored yet. Our goal thus is  to examine the differential effect of suppression of CR transport near the sources in controlled simulations 
of idealized, but realistic, galaxies representative of $\sim$\Lstar~galaxies at $z=0$ and gas-rich galaxies more typical at higher redshifts. As we show below, several qualitatively new effects emerge when CRs are allowed to accumulate near star-forming regions and retain most of their energy; in particular, the formation of massive star-forming clumps in gaseous disks becomes strongly suppressed. 

The paper is organized as follows. In Section~\ref{sec:crprop}, we discuss CR propagation models, their uncertainties, and arguments and evidence for the suppression of CR propagation near star-forming sites. In Section~\ref{sec:sims}, we discuss our simulations and implementation of different physical properties, including the treatment of CR propagation. In Section~\ref{sec:results}, we present our main results and discuss their implications for galaxy evolution in Section~\ref{sec:discussion}. We summarize our conclusions in Section~\ref{sec:summary}. In the Appendices, we present tests of the CR propagation model in the simulations, as well as additional results that are used to gauge the sensitivity of our results to variations of simulation parameters.

\section{CR propagation and possible suppression of diffusion near star-forming regions}  
\label{sec:crprop}

\subsection{Standard approaches to modeling CR transport and their limitations}

Galactic CRs are thought to be accelerated in star-forming regions predominantly by shocks around young SN remnants \citep[e.g.,][]{hillas05,Caprioli.etal.2010,Lingenfelter.2018}, with a possible contribution from stellar wind shocks  \citep[][see \citealt{Aharonian.etal.2012}, \citealt{bykov14}, and \citealt{Gabici2019} for  reviews]{Yang.etal.2018,Yang.etal.2019,Zirakashvili.Ptuskin.2018}.
The isotropy of arrival directions of the CRs detected on Earth indicates that they undergo extensive random diffusion between their injection sites and detectors.  
The average diffusion coefficient of CRs in the Milky Way is constrained to be $\kappacr \sim 10^{28}\cm2s$ at the rigidity of $\sim 1$ GV from the measurements of elemental and isotopical abundances of CR fluxes \citep[e.g.,][and references therein]{Evoli.etal.2019}.
Nevertheless, Galactic ISM is turbulent and highly inhomogeneous, and local CR diffusion can be very different from the inferred average for the Galaxy. 

Although a number of theoretical models of CR transport in different regimes have been explored extensively in the literature \citep[e.g.,][]{Zweibel.2013, Hopkins.Squire.etal.2020}, such models are highly uncertain, and there is no reliable microscale theory for macroscale CR transport coefficients yet. 
A predictive model for CR transport would need to ascertain the amplitude $\delta B(k)/B_0$ of the magnetic turbulence spectrum at the scales resonant with the CRs, which is a highly nonlinear combination of generation, damping, and (direct/inverse) cascade in $k$ space. The existing models, however, rely on rather strong assumptions about both generation and damping of resonant waves: the wave growth rates are usually computed assuming a linear theory for CR-driven streaming instability or the power spectrum of the waves excited by extrinsic turbulence, while the damping rates are typically calculated in a quasi-linear limit assuming nonlinear Landau or turbulent damping 
\citep[e.g.,][]{wiener+13,wiener+18,Zweibel.2013,aloisio+15}.

CR propagation  is commonly decomposed into diffusion and streaming terms, sometimes allowing for anisotropic diffusion. 
Modeling of streaming is based on the assumption that a steady state can be established between the growth of magnetic fluctuations due to CR-driven instabilities and some damping mechanism, with the resulting magnitude of the fluctuations setting the CR propagation velocity. 
However, because the very nature of both CR-driven field amplification \citep[e.g.,][]{kulsrud+69,skilling75a,shapiro+98,Bell.2004,amato+09} and damping \citep[e.g.,][]{kulsrud+69,mck-w69,lee+73,volk+82,goldreich+95,drury+96,farmer+04,ptuskin+06,brunetti+07,reville+08a,squire+17} are quite uncertain, a self-consistent modeling of CR diffusion coefficient is necessarily uncertain too, even if one assumes that the linear theory holds.

When the modulus of the parallel diffusion coefficient is uncertain by large factors, so is the diffusion in the direction perpendicular to the $B$ field, which is usually calculated from the former using quasi-linear theory \citep[e.g.,][]{jokipii71,matthaeus+03}.
However, when fluctuations become nonlinear at the resonant scales ($\delta B/B_0\sim 1$), the classical small-pitch-angle scattering approximation breaks down, and diffusion occurs close to the Bohm limit, in which the mean free path is comparable to the CR gyroradius \citep[e.g.,][]{reville+13,caprioli+14c}.

Moreover, complex transport models are not constrained by the existing observations of CRs in the Milky Way, as all of the main observables (CR fluxes, secondary/primary ratios, anisotropy, diffuse nonthermal backgrounds) are consistent with the isotropic diffusion of CRs \citep[e.g.,][]{strong+98,dragon13,Evoli.etal.2019}.
In addition, the modeling of both streaming and anisotropic diffusion is limited by the resolution of modern galaxy formation simulations: in highly turbulent ISM, the magnetic field is expected to be tangled on unresolved scales, which can lead to isotropic diffusive CR transport on scales resolved in simulations even if CRs were propagating exclusively along the wandering small-scale magnetic fields. 

Given these limitations, in this paper, we adopt a simple isotropic diffusion model for CR propagation and specifically focus on the effects of possible variations of CR diffusivity near the acceleration sites.

\subsection{Diffusion suppression in star-forming regions: Theory}
\label{sec:crsupp:theory}

In the vicinity of strong shocks where CRs are accelerated, each of the common assumptions used in CR propagation modeling is likely violated. This is because the CR current is much larger than in the average ISM, and nonresonant modes can be preferentially excited \citep{Bell.2004,reville+13,caprioli+14c,Caprioli.Spitkovsky.2014b}. Such plasma instabilities can amplify the background magnetic field by orders of magnitude leading to the Bohm diffusion regime \citep[e.g.,][]{blasi+07,Morlino.Caprioli.2012}. 
The CR diffusion coefficient in this regime is a factor of $\sim 10^6$ smaller than the average Galactic value at 1 GV.

Likewise, after CRs escape the accelerating region, they are expected to drive magnetic amplification via the streaming instability, which should generally result in a diffusion coefficient intermediate between the small Bohm value and the Galactic average  \citep[e.g.,][]{Yan2012,malkov+13,blasi+15,nava+16,nava+19}. Such suppression of diffusion can prolong the CR residence time in the supernova-driven bubbles well beyond the shock confinement epoch \citep{Celli2019}.

Indeed,  recent self-consistent kinetic plasma simulations predict that CR acceleration regions should be surrounded by the CR-generated ``bubbles''---underdense regions filled with magnetic turbulence that confines CRs escaping from the shock \citep{schroer+20}.
This prediction is general and does not rely on the actual regime in which magnetic field amplification occurs.
In fact, such CR bubbles are expected to expand until the CR pressure is balanced by the ambient ISM pressure, which can reach hundreds of parsecs for SN-driven superbubbles around star-forming regions. 
The exact structure and value of the corresponding  diffusion coefficient in such regions are hard to quantify, given the limited range of scales modeled in the modern kinetic simulations. Nevertheless, the results of \citet{schroer+20} strongly suggest that diffusion is indeed isotropic and occurs at a few to ten times the Bohm limit in such regions, or  several orders of magnitude below the Galactic level.

\subsection{Diffusion suppression in star-forming regions: Observations }
\label{sec:crsupp:obs}

In agreement with the emerging theoretical picture, observations also indicate that regions of active star formation are strongly correlated with nonthermal emission \citep[e.g.,][]{ackermann+11,Tabatabaei2013,Yang.etal.2018}. 
Modeling of $\gamma$-ray emission from nearby supernova remnants and molecular clouds surrounding them indicates that in these regions, the CR diffusion coefficient may be $\sim 10\text{--}100$ times smaller than the average Galactic value \citep[][]{Fujita2009,Gabici2010,Li.Chen2010,Li.Chen.2012,ajello+12,Ohira2011,uchiyama+12,Yan2012,Ohm2013, hanabata+14}. 

The diffuse $\gamma$-ray emission observed around star clusters \citep[e.g.,][]{aharonian+19} supports the idea that CR sources are surrounded by a halo of several tens of parsecs where the inferred diffusion coefficient is significantly reduced with respect to the average Galactic values, possibly by 4--5 orders of magnitude.
Previous studies showed that the degree of diffusion suppression depends on the diffusion anisotropy, but can still be significant at distances $\lesssim 50$ pc from the sources \citep{Nava.Gabici2013,nava+16,nava+19}. At the same time, recent kinetic simulations by \citet{schroer+20} suggest that a strong CR flux naturally erases the initial field geometry, eventually leading to isotropic diffusion on even larger scales, which are insensitive to the microphysics of the streaming instability and set by the CR energetics only.

Another compelling evidence of CR diffusion suppression near the sources is the presence of TeV $\gamma$-ray extended halos (tens of parsecs wide) around two nearby pulsar wind nebulae, Monogem and Geminga, recently revealed by HAWC observations \citep{hawc17}.
Such TeV halos, which have now also been discovered around other sources, are particularly intriguing because they carry pristine information about the transport of high-energy particles that are produced in the pulsar magnetosphere and accelerated in the pulsar wind, up to PeV energies in Crab-like systems.

The relatively short cooling time for inverse-Compton scattering and synchrotron emission of multi-TeV electrons ($\lesssim 10^5$~yr) allows one to constrain their diffusion time in such halos, corresponding to diffusion coefficients two to three orders of magnitude smaller than the typical Galactic one \cite[e.g.,][]{evoli+18,Bao2019}. The most plausible explanation for such suppression is again the self-confinement of  escaping particles due to exciting some kind of lepton-driven resonant instability. Although it is currently under debate whether such halos are produced directly by the relativistic leptons accelerated in the pulsar wind nebulae or by the CR protons produced at the corresponding SN shock, the increased sensitivity of the Cherenkov Telescope Array $\gamma$-ray instrument should detect many more instances of such halos and therefore constrain the size and properties of the regions where the diffusion coefficient is significantly reduced with respect to the Galactic value.

\section{Simulations} 
\label{sec:sims}

\subsection{Simulation code overview}
\label{sec:sims:overview}

We explore the effect of CR feedback and diffusivity suppression near the acceleration sites by using simulations of an isolated \Lstar~galaxy carried out with the adaptive mesh refinement $N$-body and hydrodynamics code ART \citep{Kravtsov.1999,Kravtsov.etal.2002,Rudd.etal.2008,Gnedin.Kravtsov.2011}. Our simulation setup is similar to that in \cite{Semenov.etal.2017,Semenov.etal.2018,Semenov.etal.2019}, and therefore, we only briefly describe its main features. 

The hydrodynamic fluxes in the ART code are computed using a second-order Godunov-type method \citep{Colella.Glaz.1985} with a piecewise linear reconstruction of states at the cell interfaces \citep{vanLeer.1979}. The mesh grid is adaptively refined when the gas mass in a cell exceeds $\sim8\,300 \Msun$, reaching the maximal resolution of $\Delta = 40\pc$. The gravity of gas, stars, and dark matter is solved by using a Fast Fourier Transform at the lowest grid level and relaxation method on all higher refinement levels, with the effective resolution for gravity corresponding to $\sim$2--4 cells \citep[see][]{Kravtsov.etal.1997,Mansfield.Avestruz.2020}. Radiative gas cooling and heating are modeled following \citet{Gnedin.Hollon.2012} and assuming a constant solar metallicity and a constant UV background with the H$_2$ photodissociation rate in the Lyman--Werner bands of $10^{-10}\;{\rm s^{-1}}$ \citep{Stecher.Williams.1967}. To account for the dense gas shielding from the background radiation, we use a prescription calibrated in radiative transfer simulations of the ISM \citep[the “L1a” model in][]{Safranek-Shrader.etal.2017}.

One of the key features of our simulations is the explicit dynamic modeling of unresolved turbulence following the so-called Large Eddy Simulations methodology. Our implementation in the ART code is based on the ``shear-improved’’ version of the \citet{Schmidt.etal.2014} model as detailed in \citet{Semenov.etal.2016}. In this model, the unresolved turbulent energy, $\eturb$, is sourced by the fluctuating component of the resolved velocity field that is interpreted as the onset of the turbulent cascade, and decays on the timescale close to the turbulent cell-crossing time. This unresolved turbulence provides a nonthermal pressure support to gas and facilitates diffusive turbulent transport of thermal energy and CRs (see Section~\ref{sec:sims:cr}).

The turbulence model is also directly coupled with the star formation prescription as it is used to identify the star-forming gas. Specifically, the gas is defined as star-forming when its (subgrid) virial parameter is $\avir < 10$, where the $\avir$ for simulation cells with size $\Delta$ is defined as for a uniform sphere with radius $R = \Delta/2$ \citep{Bertoldi.McKee.1992}:
\begin{equation} 
\label{eq:avir}
    \avir \equiv \frac{5 \stot^2 R}{3GM} \approx 9.35 \frac{ (\stot/10\kms)^2 }{ (n/100\cc) (\Delta/40 \pc)^2},
\end{equation}
where $\stot = \sqrt{\st^2+\cs^2}$ accounts for both the unresolved turbulent velocity dispersion, $\st = \sqrt{2\eturb/\rho}$ and thermal support. 
The local star formation rate density in such gas is parameterized as
\begin{equation} 
\label{eq:rhoSFR}
    \rhoSFR = \epsff \frac{\rho}{\tff},
\end{equation}
with a constant star formation efficiency per freefall time of $\epsff=1\%$. The choice of the constant value of $\epsff=1\%$ below and the $\avir<10$ threshold is motivated by the typical $\epsff$ and $\avir$ values estimated for the observed star-forming regions on scales comparable to our resolution \citep[e.g.,][]{Krumholz.Tan.2007,Evans.etal.2009,Evans.etal.2014,Heiderman.etal.2010,Lada.etal.2010,Lee.etal.2016,Leroy.etal.2016,Leroy.etal.2017,MivilleDeschenes.etal.2016,Utomo.etal.2017} and also approximates the exponential dependence of $\epsff$ on $\avir$ found in the MHD simulations of turbulent star-forming regions by \citet{Padoan.etal.2012,Padoan.etal.2017}.

The feedback from young stars is modeled by injecting 20\% of Type II SNe energy as CRs and the rest as the radial momentum and thermal energy computed using the fits to simulations of SN remnants evolution in a nonuniform ISM from \citet{Martizzi.etal.2015}. Our choice of $20\%$ acceleration efficiency is motivated by the fact that most of the SNe explode in the regions already populated by CRs accelerated by stellar winds and previous SNe from the same stellar population, and that rejuvenation of such preexisting CRs can lead to acceleration efficiencies significantly higher than the canonical $\sim 10\%$ \citep{Caprioli.Spitkovsky.2014a,Caprioli.etal.2018}. 
The total number of SNe for a given star particle is computed using the \citet{Chabrier.2003} IMF, and these SNe are assumed to explode uniformly in time over 3-43 Myr since the birth of the stellar particle. 
In addition to SNe, we also account for stellar mass loss and inject the mass computed from the \citet{Leitner.Kravtsov.2011} model and the corresponding linear momentum into the cell hosting the stellar particle.

Apart from CR injection, there are two differences of the stellar feedback model used in this paper from our fiducial model in \cite{Semenov.etal.2017,Semenov.etal.2018,Semenov.etal.2019}: (i) we do not boost the radial momentum of SNe and use the default \citet{Martizzi.etal.2015} values, and (ii) we use the time lag between the creation of a stellar particle and the first SN of 3 Myr. The SN momentum boosting was adopted in these papers to mitigate the loss of momentum due to the advection errors and to mimic a combined effect of clustered SNe \citep{Gentry.etal.2017,Gentry.etal.2018} and CR pressure \citep{diesing+18}. Here we turn this boosting off to demonstrate that a comparable effect can be obtained by modeling CRs with locally suppressed diffusivity.
As for (ii), the lag before the first SN does not affect the result significantly as long as this lag is shorter than the local depletion time in the star-forming gas, $\rho/\rhoSFR = \tff/\epsff \sim \text{few } 100\Myr$ for our $\epsff=0.01$ and typical freefall times in the star-forming gas of several Myr.

\def\NA{---}
\newcommand{\setcolwidth}[1]{\multicolumn{1}{>{\centering}p{0.7in}}{#1}}
\begin{deluxetable*}{lccccc}
\tablecolumns{6}
\tablecaption{Summary of the simulation parameters \label{tab:sims}}
\tablewidth{0pt}
\tablehead{
Label & 
\setcolwidth{$\fgas$\tablenotemark{a}} & 
\setcolwidth{$\zeta_{\rm cr}$\tablenotemark{b}} & 
\setcolwidth{$\kappa_{\rm cr,0}$\tablenotemark{c}} & 
\setcolwidth{$\kappa_{\rm cr,sn}$\tablenotemark{d}} & 
\setcolwidth{$n_{\rm eff,sn}/\ncell$\tablenotemark{e}}
}
\startdata
\sidehead{\it Moderate gas fraction, marginally stable disk:}
\texttt{fg0.2-noCR} & 0.2 & 0.0  & \NA & \NA & \NA \\
\texttt{fg0.2-const$\kappa$} & 0.2 & 0.2 & $10^{28}$ & $10^{28}$ & $1$ \\
\texttt{fg0.2-supp$\kappa$} & 0.2 & 0.2 & $10^{28}$ & $\sim(1\text{--}5)\times10^{25}$ & $0.01$ \\
\sidehead{\it High gas fraction, unstable disk:}
\texttt{fg0.4-noCR} & 0.4 & 0.0 & \NA & \NA & \NA \\
\texttt{fg0.4-const$\kappa$} & 0.4 & 0.2 & $10^{28}$ & $10^{28}$ & $1$ \\
\texttt{fg0.4-supp$\kappa$} & 0.4 & 0.2 & $10^{28}$ & $\sim(1\text{--}5)\times10^{25}$ & $0.01$
\enddata
\tablenotetext{a}{Gas mass fraction of the galactic disk, $\fgas = \Mg/(\Mg+M_\star)$.}
\tablenotetext{b}{Fraction of SN energy injected as CRs.}
\tablenotetext{c}{CR diffusivity in the average ISM and halo, i.e., far away from the sources (in$\cm2s$).}
\tablenotetext{d}{Effective CR diffusivity near the injection sites identified as detailed in Section~\ref{sec:sims:diffusion} (in$\cm2s$). In the runs with locally suppressed $\kappacr$, CR diffusion near the injection sites is dominated by the turbulent diffusivity that depends on the local turbulent velocity (see Equation~\ref{eq:kappa-sgst}). The values cited in the table correspond to typical $\st\sim 2\text{--}15\kms$ reached in the star-forming regions in our simulations. }
\tablenotetext{e}{Effective density for CR losses near the injection sites normalized by the cell density. Far away from the sources, $\neff=\ncell$ in all runs.}
\end{deluxetable*}

\vspace{3em}
\subsection{Modeling of Cosmic Rays}
\label{sec:sims:cr}

We model CRs as a separate fluid field by solving the advection-diffusion equation for the total CR energy density:
\begin{equation}
\label{eq:ecr}
\begin{split}
 \frac{\partial \ecr}{\partial t} &+ \nabla (u \ecr) = -\Pcr \nabla u + \nabla(\kappacr \nabla \ecr ) + \\
    & + \nabla[\rho \kappa_{\rm turb}\nabla \left({\ecr}/{\rho}\right)] -\Lambdacr + S_{\rm sn}.
    \end{split}
\end{equation}
CR advection and the $PdV$ term with $\Pcr = (\gcr - 1) \ecr$, $\gcr = 4/3$, are treated by solving the entropy conservation equation for CRs (see Appendix~\ref{app:entropy} for details). The two diffusion terms describe the isotropic CR diffusion with spatially varying $\kappacr$ (see Section~\ref{sec:sims:diffusion}) and the diffusive transport by unresolved turbulence approximated by a gradient-diffusion closure with the diffusivity 
\begin{equation} 
\label{eq:kappa-sgst}
\begin{split}
    \kappa_{\rm turb} &= \frac{c_{\kappa} }{\sqrt{2}} \st \Delta \approx\\
    &\approx 2\times10^{25} \left( \frac{\Delta}{40\pc} \right) \left( \frac{\st}{6\kms} \right) \cm2s,
\end{split}
\end{equation}
where $\st$ is modeled explicitly (see Section~\ref{sec:sims:overview}), $\Delta$ is the cell size, and $c_{\kappa}=0.4$, following \citet{Schmidt.etal.2006,Schmidt.etal.2014}. Both diffusion terms are solved using an explicit Forward Time Centered Space scheme and subcycling over the hydrodynamic step to mitigate the time-step constraint of the method (the test of the implementation is provided in Appendix~\ref{app:diffusion}). Finally, the sink and source terms associated with CR losses (Section~\ref{sec:sims:cr-losses}) and sourcing by SNe, assuming 20\% acceleration efficiency (Section~\ref{sec:sims:overview}), are added in an operator-split manner.

Our treatment of CR propagation is similar to other implementations of diffusive CR fluid in the literature \citep[e.g.,][]{Booth.etal.2013,Pfrommer.etal.2017}. 
One difference from many recent studies is that we do not model CR streaming and assume that CR diffusion is isotropic---a natural assumption for simulations without MHD. 
Although such a propagation model may appear simplistic, it is a reasonable choice given the theoretical and observational uncertainties about the CR propagation in galactic plasmas discussed in Section~\ref{sec:crprop}.
Indeed, a simple CR propagation model with isotropic diffusion and constant diffusion coefficient can account for all observations of CRs in the solar system \citep[e.g.,][]{Evoli.etal.2019}. At the same time, $\gamma$-ray observations in other galaxies are not sufficiently constraining and are consistent with a wide range of different CR propagation models, including the isotropic diffusion model with constant $\kappacr$ \citep[][]{Chan.etal.2019}, while large theoretical uncertainties do not allow a strong preference for one propagation model or its parameters over another (see Section~\ref{sec:crprop}). 

In this study, we focus on exploring the differential effect of suppression of CR transport near the sources as motivated in Section~\ref{sec:crprop}. While the overall model of CR propagation may not be accurate, the relative effect of such local suppression is interesting. Its effect is complementary to any possible variations of transport coefficients far away from CR sources, and using a simple model in the latter regime also makes the interpretation of our results more transparent.

\subsubsection{Suppression of CR diffusivity near the injection sites}
\label{sec:sims:diffusion}

Although CR diffusion in our simulations is assumed to be locally isotropic, we relax the assumption of constant diffusivity $\kappacr$ and allow it to vary spatially. We assume that CR transport in the average ISM and in the halo can be described by a constant and isotropic diffusivity $\kappa_{\rm cr,0} = 10^{28} \cm2s$ \citep[e.g.,][]{Evoli.etal.2019}, while near the sites of recent star formation, where SNe and stellar winds are expected to create and sustain low-density superbubbles, we assume that the diffusion coefficient is suppressed by a constant factor.

As detailed in Section~\ref{sec:crprop}, observations of the $\gamma$-ray emission around star clusters, isolated SN remnants, and pulsar wind nebulae suggest that this diffusion suppression factor can be rather large, up to several orders of magnitude. 
To bracket the possible range of suppression, we reduce $\kappacr$ in the regions with active SNe such that the CR diffusion becomes dominated by the turbulent advection by unresolved eddies, which is the lowest limit for the CR diffusivity in our simulations (the second diffusive term in Equation~\ref{eq:ecr}). From Equation~(\ref{eq:kappa-sgst}), using the typical values of $\st\sim 2\text{--}15\kms$ reached in the star-forming regions in our simulations, $\kappa_{\rm turb}$ is 200--1000 times smaller than $\kappa_{\rm cr,0}$, which is within the range between the diffusion suppression factors expected for the shock vicinity (i.e., Bohm diffusion with $\kappacr \sim 10^{-6}\; \kappa_{\rm cr,0}$) and the values inferred for the extended regions surrounding CR sources ($\kappacr \sim 0.01\text{--}0.1\;\kappa_{\rm cr,0}$; see Section~\ref{sec:crprop} and references therein).
To separate the effects of local CR diffusion suppression from other effects of CR feedback, we have also run a model, in which the diffusion coefficient is constant in space with the value of $\kappacr = 10^{28}\cm2s$. In Appendix~\ref{app:var-factors}, we also demonstrate the sensitivity of our results to variation of the diffusion suppression factor.

To identify the cells in the simulation where CR diffusion is to be suppressed, we introduce a passively advected scalar field that counts down the time since the most recent star formation event, $t_{\rm age}$. To this end, at each time step and in each cell, $t_{\rm age}$ is set to the age of the youngest stellar particle within the cell and the size of the time step is subtracted from $t_{\rm age}$. Having the spatial distribution of $t_{\rm age}$, we suppress $\kappacr$ in the cells with $t_{\rm age}<t_{\rm SN}=40\Myr$, gas density $n>1\cc$, and temperature $T<10^5\K$. The $t_{\rm age}$ cut selects the gas in the vicinity of recent star formation and SN activity, and the choice of the $t_{\rm SN}=40\Myr$ threshold is motivated by the timescale over which shocks and superbubbles can be sustained by repeating SN explosions in a given single-age population of stars. 
The additional density and temperature cuts prevent the suppression of diffusion inside the resolved hot SN bubbles: $\kappacr$ is expected to be suppressed only upstream of the shock while in the bubble interior it can become large again. We find that not applying these additional cuts results in significantly lower CR energy density inside the SN bubbles due to the advection of CRs by expanding SN shells, but it has little effect on the ISM density structure and SFR.

It is worth noting that for a given $t_{\rm SN}$, the effect of CR diffusivity suppression saturates at sufficiently large values of the suppression factor. The saturation happens when the escape time of CRs from the regions with a suppressed $\kappacr$ is longer than the duration of the suppression, $t_{\rm SN}$:
\begin{equation}
\label{eq:diff-time}
    t_{\rm diff} \sim \frac{l^2}{\kappacr} \sim 0.3\;l_2^2\;\kappa_{28}^{-1} \Myr > t_{\rm SN},
\end{equation}
where $l_2 \equiv l/100\pc$ and $\kappa_{28}  \equiv \kappacr/10^{28}\cm2s$. For $t_{\rm SN}=40\Myr$, this condition implies that as long as the reduced diffusivity is $\kappacr<10^{26}\cm2s$, the CR residence time is limited by the timescale on which SN bubbles are sustained on unresolved scales. The turbulent diffusivity $\kappa_{\rm turb}$ is below this critical value, and we indeed see no significant effect on our results when we set $\kappa_{\rm turb} = 0$, effectively switching off CR diffusion in regions with suppressed $\kappacr$ (see Appendix~\ref{app:var-factors}).

In the ART code, tracking the time since the most recent SF event is also used in the implementation of the so-called ``blastwave'' or ``delayed cooling'' feedback model to identify the regions where the radiative cooling is suppressed after SN explosions \citep[][see also Section~\ref{sec:discussion} for the further comparison with the delayed cooling feedback]{Gnedin.2014}. 
Note, however, that such an implementation of CR diffusion suppression is only suitable when SN-driven shocks are not resolved, as is the case for our resolution of $\Delta = 40\pc$. In this case, the CR diffusion coefficient can be suppressed in the cells with active SNe, which will also contain the upstream regions of the shocks where CR diffusion is expected to be suppressed. At higher resolution, when SN shocks become resolved, a more refined model should be used that would identify such shocks on the fly and suppress CR diffusion in their upstream regions.

\vspace{2em}
\subsubsection{CR losses and heating}
\label{sec:sims:cr-losses}

To model CR losses and heating, we adopt the rate coefficients from \citet{Pfrommer.etal.2017}. Specifically, the CR energy losses, $\Lambdacr$ in Equation~(\ref{eq:ecr}), are parameterized as 
\begin{equation}
    \Lambda_{\rm cr} = \lambda_{\rm cr}\, \neff\, \ecr,
\end{equation}
where $\lambda_{\rm cr} = 1.022\times10^{-15}\ccs$ and $\neff$ is the effective gas density for CR losses. Assuming that all Coulomb and 1/6 of hadronic losses are thermalized, a corresponding source term is added in the equation for the thermal energy:
\begin{equation}
    \Gamma_{\rm th} = \lambda_{\rm th}\, \neff\, \ecr,
\end{equation}
with $\lambda_{\rm th} = 4.02\times10^{-16}\ccs$. 
These values of $\lambda_{\rm cr}$ and $\lambda_{\rm th}$ are derived for a fully ionized medium but they can mildly change in the neutral medium as Coulomb losses decrease while ionization losses become important. However, the resulting effect on the net CR losses and gas heating is expected to be small because, for $\sim1$ GeV protons, (i) net losses are dominated by hadronic interactions and (ii) Coulomb losses in a fully ionized medium are comparable to ionization losses in a neutral medium \citep[e.g.,][]{Schlickeiser}, and therefore, we ignore the dependence of $\lambda_{\rm cr}$ and $\lambda_{\rm th}$ on the ionization state of the gas.

The rate of CR losses and heating is proportional to the effective ambient gas density, $\neff$, which depends on the complex structure of recent star formation sites where winds and photoionization from massive stars and their subsequent explosions as SNe are expected to create multiphase, low-density superbubbles. At the grid scale of our simulations or resolution of any cosmological simulations of galaxy formation, this complex gas structure is not resolved, and thus, we cannot simply use the average gas density in a grid cell, $n_{\rm cell}$, as $\neff$. 

The choice of $\neff$ is particularly important in the regions where CR diffusivity is suppressed. The physical picture motivating such suppression (Section~\ref{sec:crprop}) implies that a significant fraction of CRs are ``locked'' inside the tenuous SN bubbles \citep[see also][]{schroer+20}, and therefore, assuming $\neff = n_{\rm cell}$ would grossly overestimate the CR losses.

Ideally, $\neff$ must be predicted by a subgrid model. For example, with a model for the structure of gas and CRs on unresolved scales and assuming a constant CR spectrum (and thus $\lambda_{\rm cr}=\const$), $\neff$ could be computed as the average gas density weighted by CR energy density. However, in the absence of such a model, in this study, we simply parameterize the unresolved density structure in the star-forming sites with suppressed CR diffusion  as $\neff = \floss n_{\rm cell} < n_{\rm cell}$, and use $\neff = n_{\rm cell}$ outside of such regions. The specific value of $\floss$, is a free parameter of the model, and in the paper, we use $\floss = 10^{-2}$ motivated by the expected low densities in SN superbubbles. In Appendix~\ref{app:var-factors}, we also demonstrate the sensitivity of our results to variation of $\floss$.

It is worth noting that similarly to the diffusion suppression (see Equation~\ref{eq:diff-time}), the effect of reduced $\neff$ also saturates at small $\floss$ because the cooling time can become longer than the duration of suppression:
\begin{equation}
\label{eq:loss-time}
    t_{\rm loss} \sim \frac{1}{\lambda_{\rm cr} \neff} \sim 0.3\; (\lambda_{15}\; \floss\; n_{\rm cell,2})^{-1} \Myr > t_{\rm SN},
\end{equation}
where $\lambda_{15} \equiv \lambda_{\rm cr}/10^{-15}\ccs$ and $n_{\rm cell,2} \equiv n_{\rm cell}/100\cc$. For $t_{\rm SN}=40\Myr$ and typical average densities of star-forming cells, $n_{\rm cell,2} \sim 30\text{--}100\cc$, the effect of $\floss$ saturates for $\floss<0.01\text{--}0.03$. This saturation is demonstrated in Appendix~\ref{app:var-factors}.

\subsection{Summary of the runs and the galaxy model}
\label{sec:sims:galmodel}

\begin{figure*}
\includegraphics[width=\textwidth]{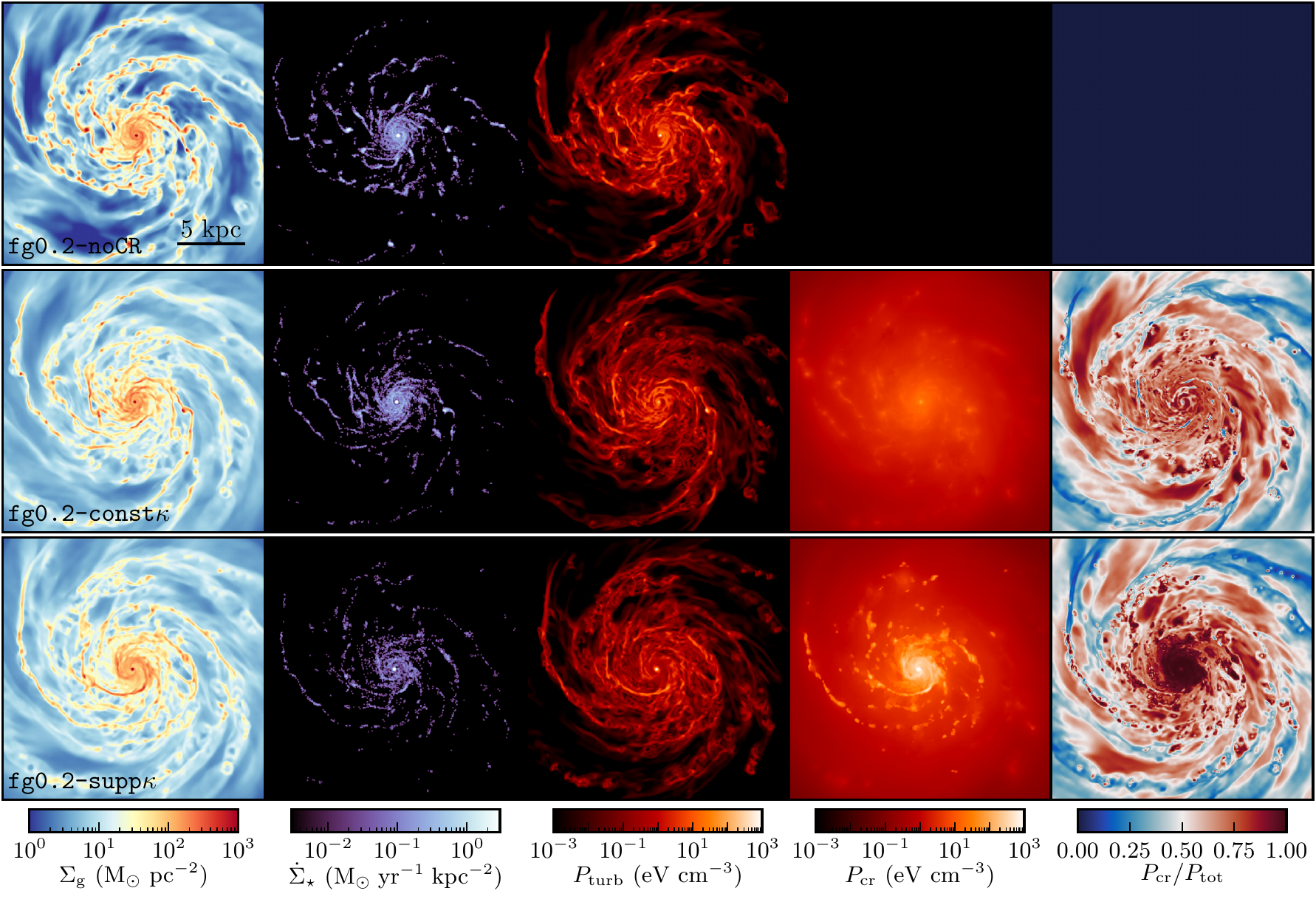}
\caption{\label{fig:faceon-fg0.2} Face-on maps of gas surface density, SFR, turbulent pressure, CR pressure, and CR pressure fraction at the disk midplane. The rows from top to bottom show runs without CR feedback (\texttt{fg0.2-noCR}), with CR feedback and constant diffusivity $\kappacr=10^{28}\cm2s$ (\texttt{fg0.2-const$\kappa$}), and with $\kappacr$ suppressed in star-forming regions, as motivated in Section~\ref{sec:crprop} (\texttt{fg0.2-supp$\kappa$}). The snapshots are shown at $t=600\Myr$ from the start of the simulation. The SFR surface density is measured using particles with ages $<$30 Myr. The CR feedback makes the galactic disk more stable and less susceptible to clump formation, especially when the CR diffusivity is suppressed in star-forming regions.}
\end{figure*}

To separate the effects of CR feedback and CR diffusivity suppression, we run our simulations in three regimes: 
(i) no CRs, with all SN feedback injected as radial momentum and heat, (ii) CR feedback with a constant diffusivity of $\kappacr=10^{28}\cm2s$, and (iii) $\kappacr$ suppressed in star-forming regions as described above. Table~\ref{tab:sims} summarizes the simulation parameters used in this paper.  

For a galaxy model, we use the initial conditions from the AGORA code comparison project \citep{Kim.etal.2014,agora2}. It is an \Lstar~galaxy with an exponential stellar and gaseous disk with a scale radius of $\approx 3.4\kpc$, scale height of $\approx 340\pc$, total mass of $\Mg+M_\star \approx 4.3\times 10^{10} \Msun$, and gas fraction of $\fgas=\Mg/(\Mg+M_\star) \approx 20\%$. The galaxy has a stellar bulge with a total mass of $\approx 4.3\times10^9 \Msun$ and a \citet{Hernquist.1990} density profile with the scale radius of $\approx 340\pc$, and it is embedded in a dark matter halo with a Navarro--Frenk--White profile \citep{NFW.1996,NFW.1997} with the total mass of $M_{\rm 200} \approx 1.074\times10^{12}\Msun$ and the concentration of $c_{200}=10$ within the radius enclosing density contrast of 200 relative to the critical density at $z=0$.

Over the initial $\lesssim\,200$ Myr of evolution, this galaxy undergoes a transient stage as it settles down. In order to mitigate the effect of this relaxation stage on our results, we restart our runs with different CR feedback models from the same simulation output, saved after this transient stage had passed. Specifically, we use a snapshot at $t=300\Myr$ from a simulation without CR feedback and with SN momentum boosted by a factor of 5, and start our simulations with the boosting of SN momentum turned off and CR feedback turned on. The changes in the feedback prescription lead to the second transient stage. For this reason, we analyze snapshots after these transient effects disappear, at $t = 600\Myr$ from the start of the simulation.

Apart from the model with a gas fraction of $\fgas\sim20\%$, we also explore the case of high gas fraction, $\fgas\sim40\%$, which is more typical for galaxies around the peak of cosmic star formation. A galactic disk with such a high gas fraction becomes gravitationally unstable, and therefore, it has to be set up particularly carefully. To this end, we restart from the $t \sim 600 \Myr$ snapshots from our $\fgas\sim20\%$ simulations and gradually increase the gas mass of the galaxy until $\fgas$ reaches $\sim40\%$ at $t\sim 650\Myr$. Although we increase $\fgas$ manually by gradually increasing gas densities in all cells within the disk, such an increase of $\fgas$ can also mimic rapid gas accretion taking place in real galaxies at high redshifts.

In the analysis presented below, we remove the region within $R<1\kpc$ from the disk center. The ISM structure near the centers of \Lstar~galaxies can be strongly affected by the AGN feedback, which is not modeled in our simulations. By removing the disk center, we highlight the effect of CR feedback on the ISM structure in the average disk, where the effects of AGN feedback are expected to be less important.

\section{Results} 
\label{sec:results}

\subsection{Effect on the galaxy structure} 
\label{sec:results:maps}

\begin{figure}
\includegraphics[width=\columnwidth]{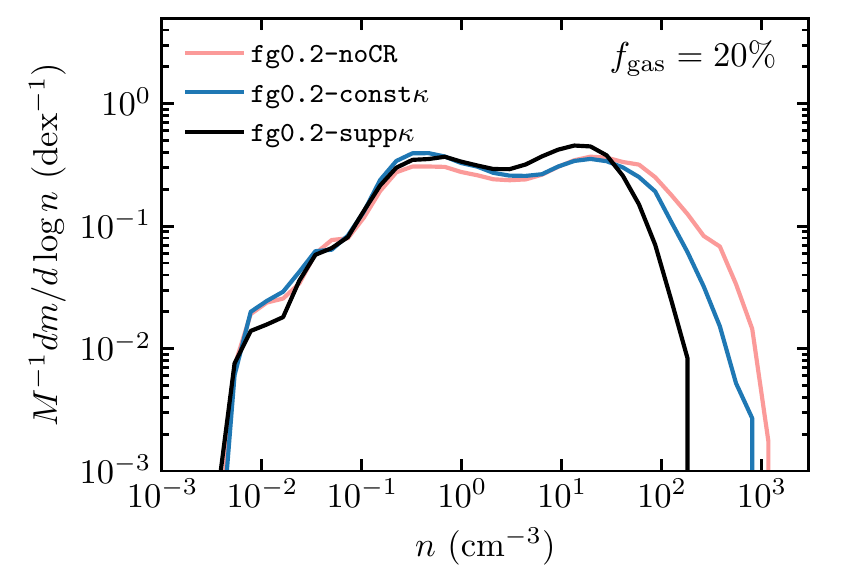}
\caption{\label{fig:pdf-fg0.2} Mass-weighted density PDFs at galactocentric radii of $R=1\text{--}20\kpc$ and within $|z| < 300\pc$ from the midplane. To reduce the noise due to the snapshot-to-snapshot variation, we show the median PDFs calculated using 11 snapshots between 500 and 600 Myr. CR feedback with constant diffusivity (\texttt{fg0.2-const$\kappa$}) reduces the amount of dense gas, but the highest densities reached in the simulation remain qualitatively similar to the run without CRs (\texttt{fg0.2-noCR}). In contrast, suppression of CR diffusivity in star-forming regions (\texttt{fg0.2-supp$\kappa$}) eliminates high-density clumps and reduces the maximal densities reached in the disk by a factor of $\sim 5$.}
\end{figure}

To gauge the effects of CR diffusion suppression on the galaxy evolution, we first compare the results of simulations of the \Lstar~galaxy with a moderate gas fraction, $\fgas=\Mg/(\Mg+M_\star)\sim20\%$, resimulated with and without CR feedback and with and without CR diffusion suppression near the injection sites.

The effect of CR feedback and diffusion suppression is readily apparent in the face-on images of different quantities of the simulated galactic disks shown in Figure~\ref{fig:faceon-fg0.2}. In the runs with CR feedback, the gas and SFR distributions (first and second columns, respectively) become smoother, and the number of dense star-forming clumps is significantly reduced. The suppression of CR diffusion around star-forming regions makes this effect even stronger. In particular, the ISM becomes completely devoid of such clumps, and gas and young stars form pronounced spiral arms reminiscent of the observed grand-design spiral galaxies.

CR feedback makes the galactic disk more stable and less susceptible to clump formation by contributing to the pressure support of the gas. Without CRs, the gas is supported only by thermal and turbulent pressure, with thermal pressure dominating in the diffuse interarm regions and turbulent pressure supporting dense and cold regions (see the middle column in Figure~\ref{fig:faceon-fg0.2}). Adding CR feedback with constant diffusivity introduces an additional smooth pressure component that becomes dominant in the volume-filling diffuse gas and therefore improves the overall stability of the disk. Although the formation of dense gas in such a disk is slowed down, some of the clumps are still able to form because CRs quickly escape from such a region. In contrast, when CR diffusivity is suppressed near the injection sites, CRs start accumulating in the dense gas and thereby create strong local pressure gradients that counteract gas compression and prevent the formation of dense clumps.

To highlight this effect of CR diffusivity suppression, Figure~\ref{fig:pdf-fg0.2} compares the probability density functions (PDFs) of gas density in these three runs. CRs with constant diffusivity do reduce the amount of dense gas but the highest densities reached in the simulations with and without CRs are nevertheless similar, $n\sim1000\cc$. On the other hand, suppression of CR diffusivity in star-forming regions strongly reduces the high-density tail of the PDF, and the highest densities reached in the simulation drop by a factor of $\sim$5. This effect is qualitatively similar to the effect of high local star formation efficiency, $\epsff$, coupled with a model of efficient local feedback that also enhances feedback and suppresses dense gas formation \citep[e.g.,][]{Orr.etal.2017,Semenov.etal.2018,Smith.etal.2020}.

\begin{figure*}
\includegraphics[width=\textwidth]{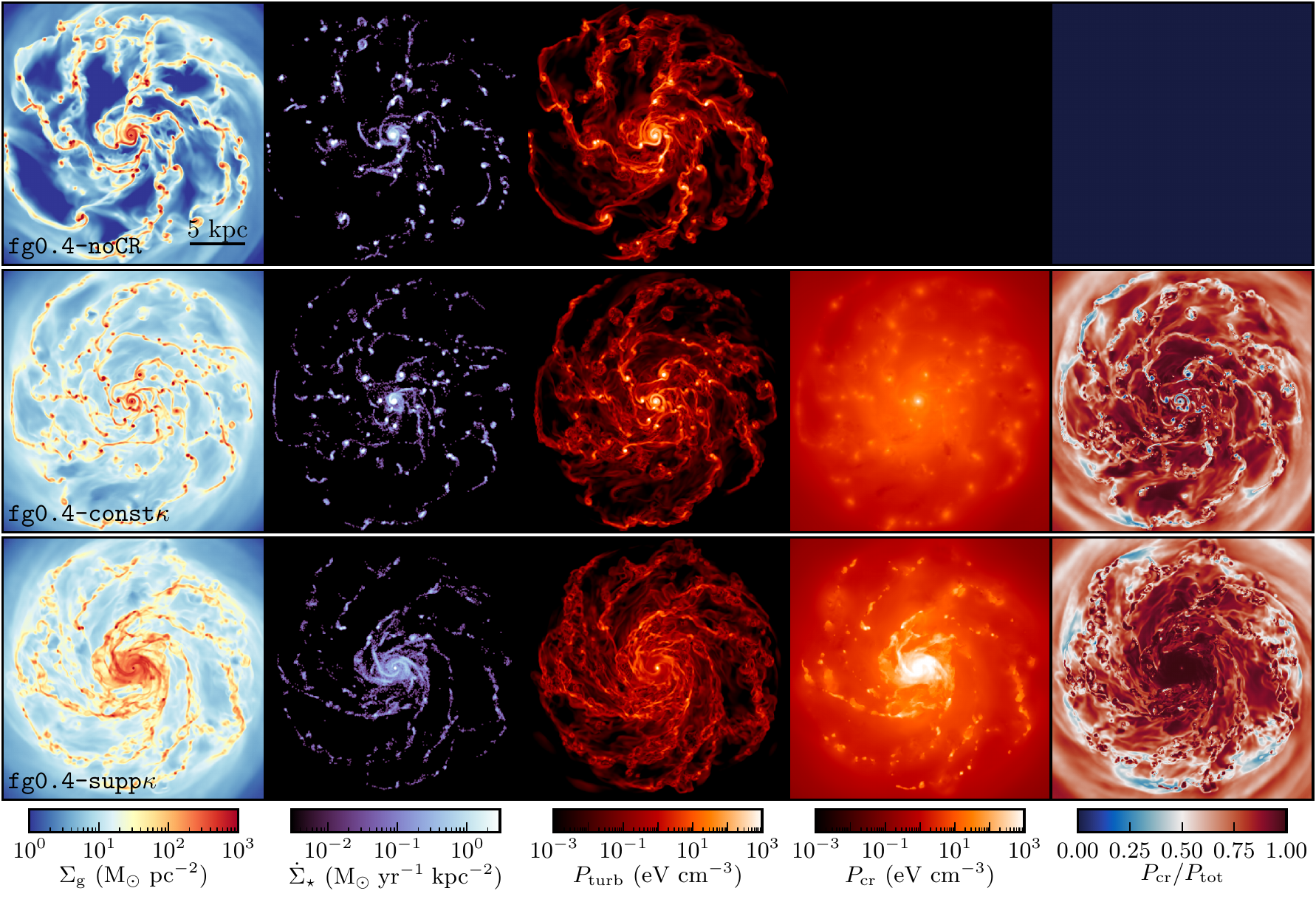}
\caption{\label{fig:faceon-fg0.4} Same as Figure~\ref{fig:faceon-fg0.2} but for the high $\fgas\sim40\%$ disk at $t=800\Myr$. A higher gas fraction results in a significantly more unstable disk, and the stabilizing effects of CR diffusivity suppression become even more apparent. In the runs without CRs (\texttt{fg0.4-noCR}) or with constant $\kappacr$ (\texttt{fg0.4-const$\kappa$}), the disk fragments into long-lived star-forming gas clumps. In contrast, suppression of $\kappacr$ near the injection sites (\texttt{fg0.4-supp$\kappa$}) prevents clump formation, and the disk can maintain a regular spiral structure. }
\end{figure*}

\begin{figure}
\includegraphics[width=\columnwidth]{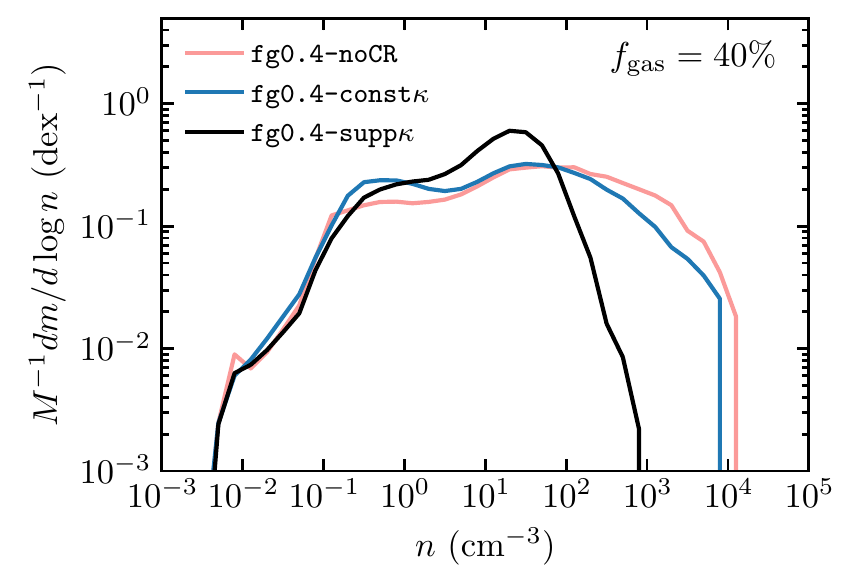}
\caption{\label{fig:pdf-fg0.4} Same as Figure~\ref{fig:pdf-fg0.2} but for the high $\fgas\sim40\%$ disk. The PDFs are stacked over 11 snapshots between 700 and 800 Myr, with lines showing the median. The effect of CR feedback on the density PDF remains qualitatively the same, but its magnitude is larger. Note that the range of densities in the horizontal axis is larger than in Figure~\ref{fig:pdf-fg0.2}. }
\end{figure}

Figures~\ref{fig:faceon-fg0.4} and \ref{fig:pdf-fg0.4} show the results for the simulations of a galaxy with the gas fraction of $\fgas\sim 40\%$. The changes in the ISM structure and pressure support introduced by CR feedback are qualitatively similar to the $\fgas\sim20\%$ case, but their magnitude is more dramatic and the contribution of CR pressure becomes dominant throughout the disk (see Appendix~\ref{app:pressure} for a detailed comparison of midplane pressure profiles). Without CRs, the entire disk fragments into long-lived star-forming clumps. CRs with constant diffusivity hinder this fragmentation somewhat but still cannot prevent it completely. In contrast, when CR diffusion is suppressed near the injection sites, CR pressure gradients become sufficiently strong so that the gaseous disk can maintain a regular spiral structure and avoid fragmentation, thereby qualitatively changing the morphology of the disk. Note that dense clumps are still able to form in such a disk, but these clumps are short lived as they are quickly dispersed by a combined effect of local CR pressure gradients and the momentum injected by SNe.

\begin{figure*}
\centering
{\large Moderate gas fraction galaxy, $f_{\rm gas} = 20\%$:}\\
\includegraphics[width=\textwidth]{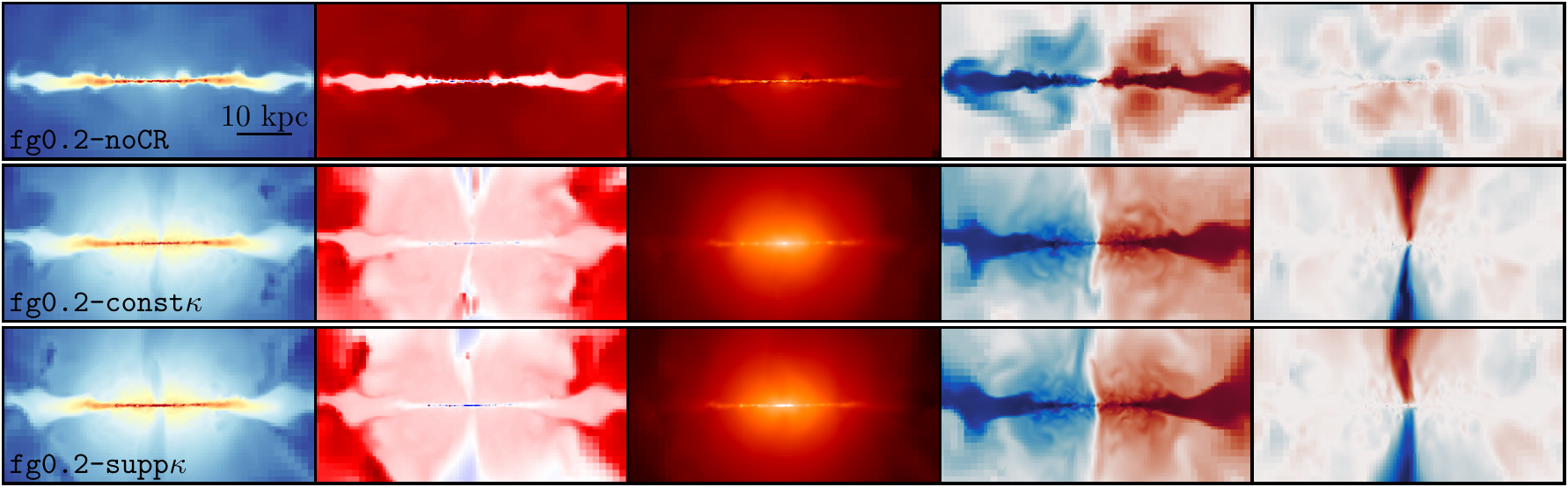}\\
\vspace{5pt}
{\large High gas fraction galaxy, $f_{\rm gas} = 40\%$:}\\
\includegraphics[width=\textwidth]{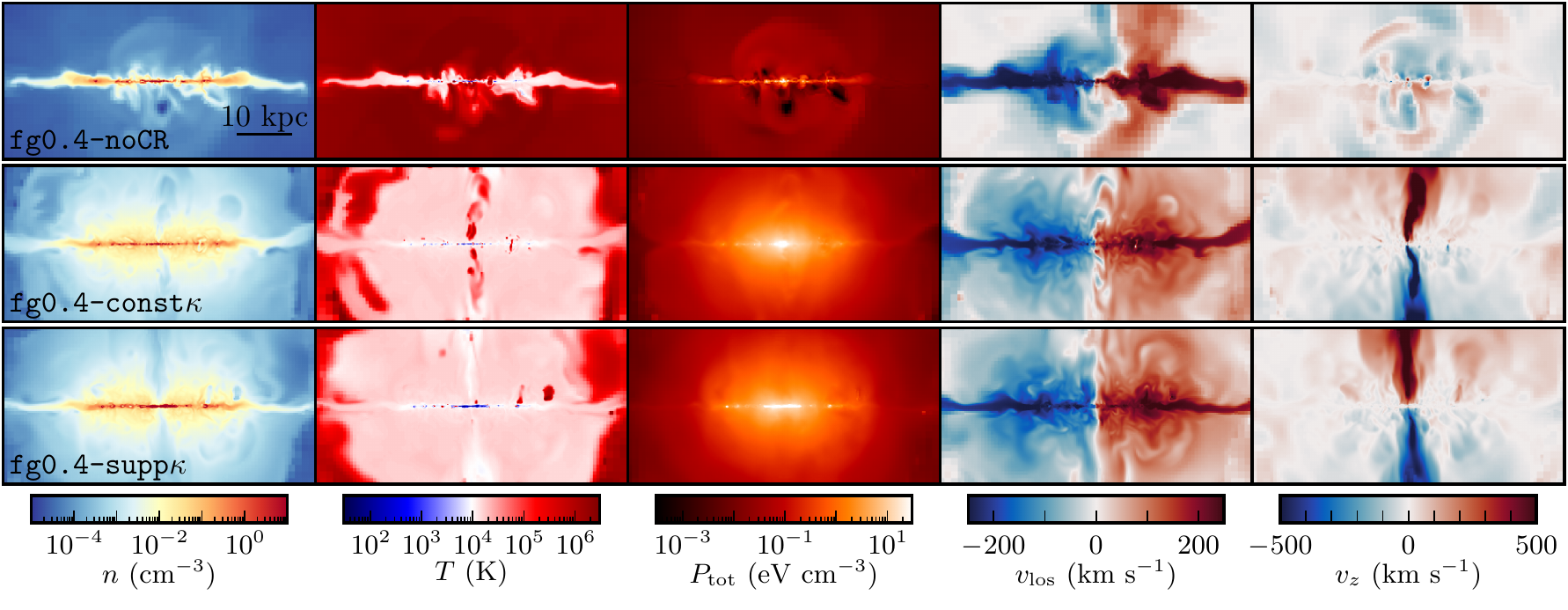}
\caption{\label{fig:edgeon} Edge-on view of the simulated galaxy with $\fgas\sim20\%$ (top set of panels) and $\fgas\sim40\%$ (bottom set of panels) for different CR feedback treatments. Columns from left to right show slices of density, temperature, total (thermal+turbulent+CR) pressure, line-of-sight velocity, and vertical velocity. The maps are shown at the same times as those in Figures~\ref{fig:faceon-fg0.2} and \ref{fig:faceon-fg0.4}: $t=600\Myr$ for $\fgas = 20\%$ and $t=800\Myr$ for $\fgas = 40\%$. The quickly diffusing component of CRs establishes an extended vertical pressure gradient that lifts some of the ISM gas and creates a tenuous warm halo corotating with the disk. Although the outflow velocities near the center can reach $\sim 500\kms$, the total mass-loading factor of this outflow is rather small: $\eta = \dot{M}_{\rm out}/\SFR \sim 0.2$ in the run with $\fgas\sim20\%$ and uniform $\kappacr$ and $\eta \sim 1$ for $\fgas\sim40\%$ with locally suppressed $\kappacr$. The vertical distribution of gas is qualitatively similar in the runs with different gas fractions and only weakly sensitive to the local suppression of CR diffusivity. }
\end{figure*}

Apart from the effect on the ISM, CR feedback also alters the distribution of gas in the halo next to the disk, as shown in Figure~\ref{fig:edgeon}. Quickly escaping CRs establish an extended vertical pressure gradient that lifts some of the gas from the ISM into a tenuous warm halo with $n\sim10^{-3}$ to a few $0.01\cc$ and $T \sim 2\times 10^4\K$ that corotates with the disk at a velocity decreasing with height. Most of the gas in this halo has a small outward velocity of a few tens of $\kms$, except for the central region where gas can be accelerated to $\sim 500 \kms$. The total mass loading of this outflow is rather small: $\eta = \dot{M}_{\rm out}/\SFR \sim 0.2$ in the run with $\fgas\sim20\%$ and uniform $\kappacr$ and increasing to $\eta \sim 1$ for $\fgas\sim40\%$ with locally suppressed $\kappacr$. To compute $\eta$, we measure the outflow rate, $\dot{M}_{\rm out}$, as a net mass flux through circular horizontal surfaces with $R<20\kpc$ at different heights $z$ above the disk and cite the maximal value of $\eta$ that in our galaxies is reached at $z\sim1\text{--}3\kpc$. 

As the figure shows, the results remain qualitatively similar for the runs with a higher gas fraction. However, the previous studies \citep[e.g.,][]{Uhlig.etal.2012,Booth.etal.2013} showed that CRs drive wind efficiently in lower-mass dwarf galaxies, and our preliminary results using the same model in simulations of dwarf-like galaxies (not presented here) confirm this. 

Another remarkable conclusion from Figure~\ref{fig:edgeon} is that the distribution of gas in the halo is insensitive to the local suppression of CR diffusivity near the injection sites. In our model, the local CR diffusion is suppressed only in the vicinity of SN II activity, so that after the last SN explodes, the CR diffusion coefficient returns to its galactic value, $\kappacr  = 10^{28}\cm2s$, and previously accumulated CRs quickly escape into the ISM and adjacent halo. Thus, local suppression of diffusion can be thought of as a delayed release of CRs from the star-forming regions, and the morphology of the diffuse CR pressure component, and therefore, the midplane pressure and vertical pressure gradient are only weakly sensitive to the suppression of $\kappacr$.

\subsection{Effect on the star formation rates} 
\label{sec:results:sfr}

\begin{figure}
\includegraphics[width=\columnwidth]{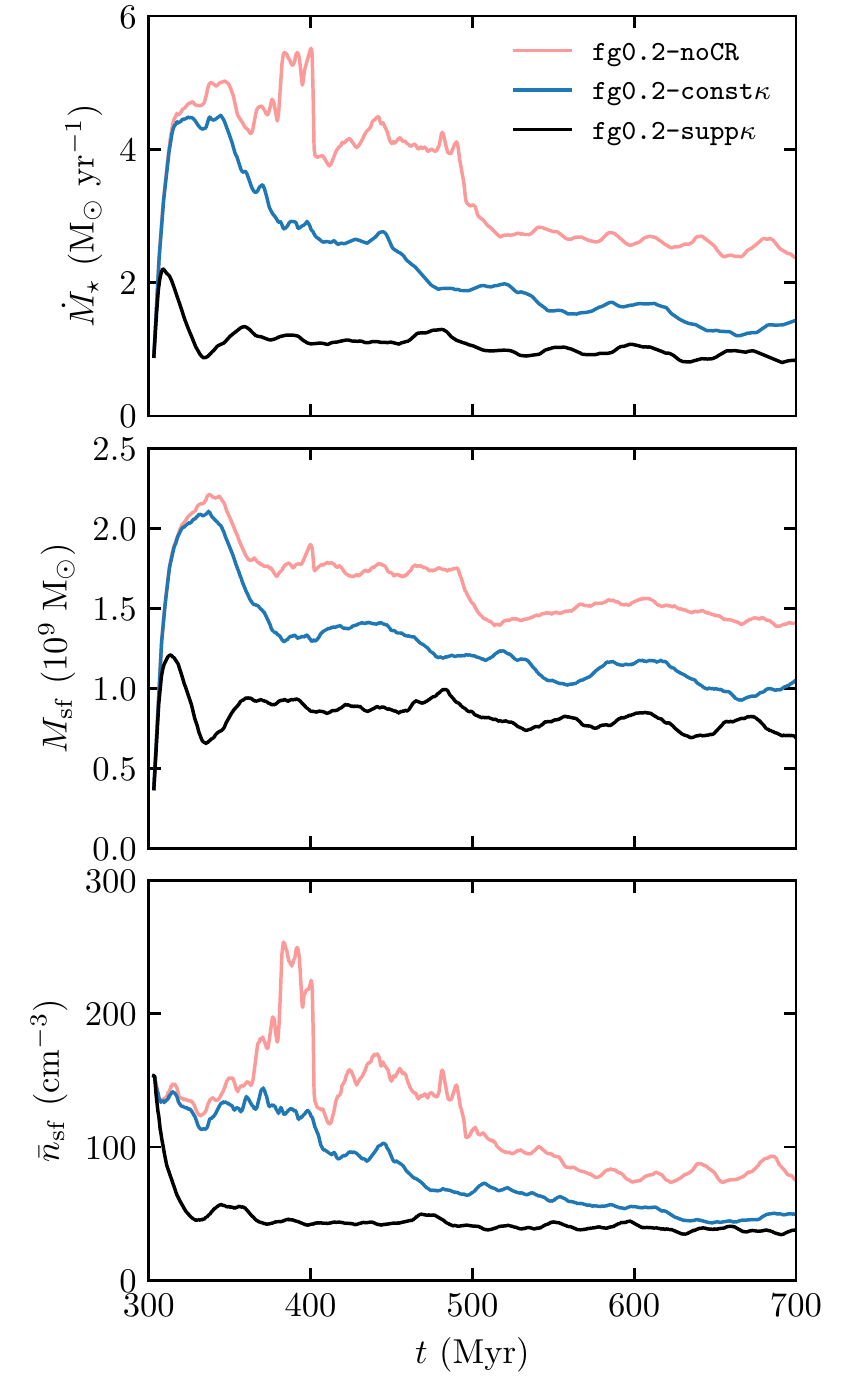}
\caption{\label{fig:sfh-fg0.2} Effect of CR feedback on the global SFR (top panel) and on the total mass, $\Msf$ (middle panel), and average density, $\bar{n}_{\rm sf}$ (bottom panel) of star-forming gas (as defined in the text following Equation~\ref{eq:sfr}). CR feedback with suppressed diffusivity (\texttt{fg0.2-supp$\kappa$}) reduces the SFR by a factor of 3--4 compared to the run without CRs (\texttt{fg0.2-noCR}). This effect is mostly due to a decrease of $\Msf$, but at early times, it is enhanced by the difference in $\bar{n}_{\rm sf}$. To highlight the effect in the average ISM, we remove the disk center, $R<1\kpc$, from the analysis.}
\end{figure}

Results presented in the previous section indicate that CR feedback can stabilize galactic disks and slow down the formation of dense gas, especially in gas-rich galaxies. This effect can naturally lead to suppression of the star formation rate (SFR) due to a combination of two factors: (i) the amount of dense star-forming gas decreases and (ii) the average density of star-forming gas is lower and thus its freefall time is longer. To gauge the relative importance of these two factors, we express the global SFR as
\begin{equation}
\label{eq:sfr}
\begin{split}
\SFR &= \int \rhoSFR dV = \int_{\rm sf} \frac{\epsff \rho}{\tff} dV =\\
     &=\epsff \Msf \left\langle \frac{1}{\tff} \right\rangle_{\rm sf} \propto \epsff \Msf \bar{n}_{\rm sf}^{0.5},
\end{split}
\end{equation}
where $\epsff=0.01$ is the local star formation efficiency per freefall time that is assumed constant, and $\Msf$ and $\bar{n}_{\rm sf}$ are the total mass and the appropriately weighted average density of the star-forming gas, respectively. The latter is computed from $\langle 1/\tff \rangle_{\rm sf}^{-1} \equiv \sqrt{ 3\pi / (32 G \mu m_{\rm p} \bar{n}_{\rm sf} )}$, assuming $\mu=1$ and where $\langle ... \rangle_{\rm sf}$ denotes the mass-weighted average over the star-forming regions. 

The evolution of $\SFR$, $\Msf$, and $\bar{n}_{\rm sf}$ in our simulated galaxy with $\fgas\sim20\%$ is shown in Figure~\ref{fig:sfh-fg0.2}. As the top panel shows, the CR feedback with locally suppressed diffusivity reduces the SFR by a factor of 3--4 compared to the run without CRs. For comparison, such a suppression of SFR is comparable to the effect of SN momentum boosting by a factor of $b\sim 5\text{--}7$ \citep[in][we find $\SFR \propto b^{-0.75}$ for our simulated galaxy]{Semenov.etal.2018}. 
As the bottom two panels show, at $t \lesssim 500\Myr$, the difference in SFRs results from both larger $\Msf$ and higher $\bar{n}_{\rm sf}$ in the runs without CRs and with constant $\kappacr$, while at later times, $\bar{n}_{\rm sf}$ in these two runs decreases, and the difference in the SFR becomes mainly due to the difference in $\Msf$. This decrease of $\bar{n}_{\rm sf}$ reflects a gradual reduction of the dense clump formation caused by a decrease of gas surface densities due to the global gas consumption.
We explore the sensitivity of these results to the degree of CR diffusivity suppression and the choice of effective densities for CR losses in  Appendix~\ref{app:var-factors}.  

\begin{figure}
\includegraphics[width=\columnwidth]{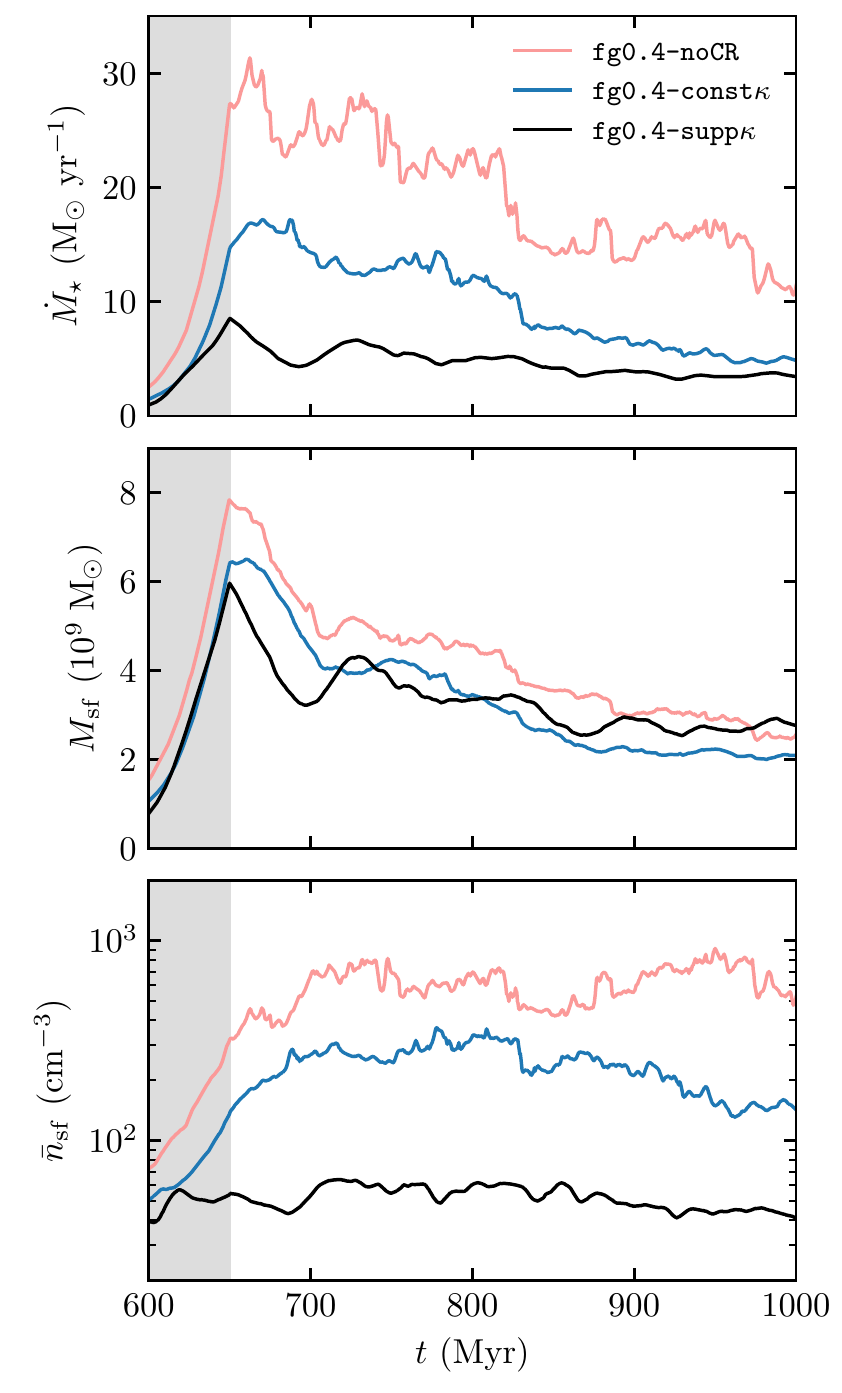}
\caption{\label{fig:sfh-fg0.4} Same as Figure~\ref{fig:sfh-fg0.2} but for the simulation with $\fgas\sim40\%$. The shaded region shows the initial phase where the gas fraction is manually increased from $\sim$20\% to $\sim$40\% as detailed in Section~\ref{sec:sims:galmodel}. After this increase of $\fgas$, the effect of CRs on SFR is somewhat stronger but qualitatively similar to the simulation with $\fgas\sim20\%$. However, this effect on the SFR is almost entirely due to the high average densities of star-forming clumps (bottom panel), while the differences in the total mass of star-forming gas among the three runs are small (middle panel). }
\end{figure}

As Figure~\ref{fig:sfh-fg0.4} shows, the effect on the SFR in the high-$\fgas$ galaxy is somewhat stronger but qualitatively similar to that in the run with $\fgas \sim 20\%$. One interesting difference from the $\fgas \sim 20\%$ case is that the effect is caused mainly by the difference in the densities of star-forming gas, while the total amount of such gas is similar in all three runs. 
As the bottom panel shows, during the initial ``accretion’’ phase when $\fgas$ is manually increased, the $\bar{n}_{\rm sf}$ increases by a factor of  $\sim$10 and $\sim$5 in the runs without CRs and with constant $\kappacr$, respectively. Such a strong increase is due to the rapid disk fragmentation and formation of numerous massive clumps. In contrast, when the local CR diffusivity is suppressed, gas compression is locally halted, and $\bar{n}_{\rm sf}$ stays at approximately the same value. As a result, although all three runs contain approximately the same amount of star-forming gas, in the run with locally suppressed $\kappacr$ this gas is smoothly distributed in the spiral arms and forms stars at significantly slower rates (recall Figure~\ref{fig:faceon-fg0.4} for a visual impression).

\begin{figure}
\centering
\includegraphics[width=\columnwidth]{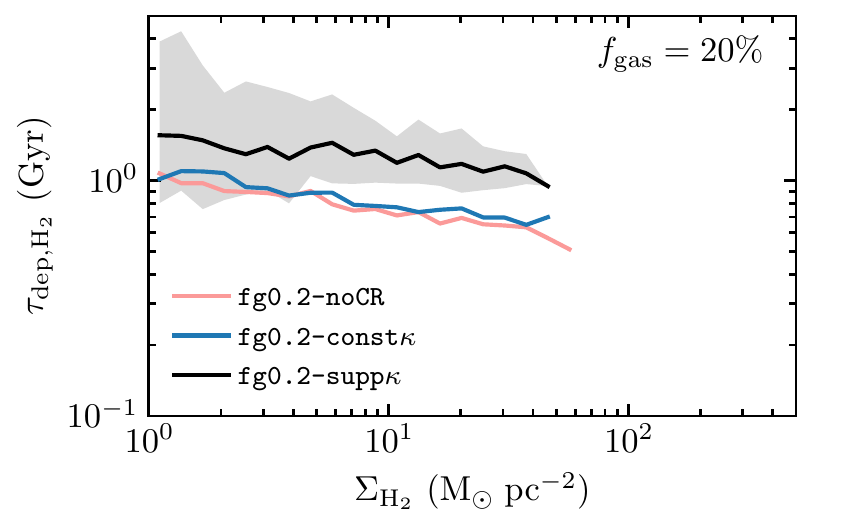}\\
\includegraphics[width=\columnwidth]{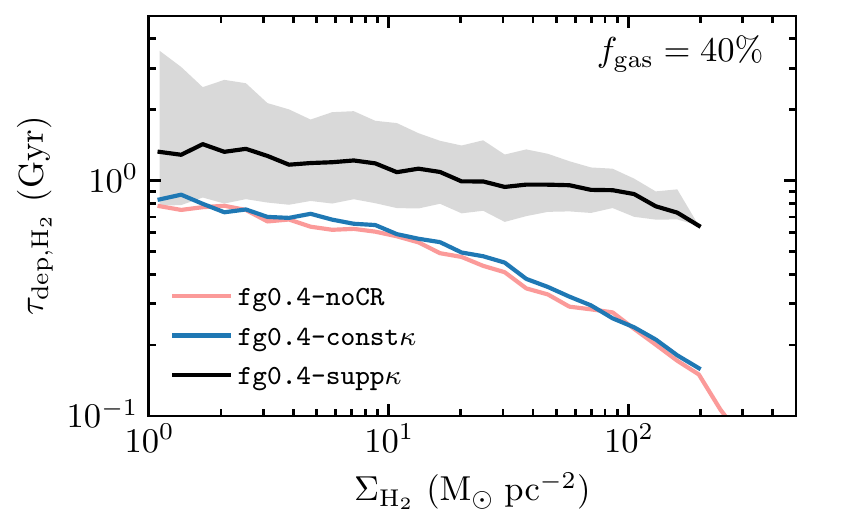}
\caption{\label{fig:tauH2} Effect of CR diffusivity suppression on the molecular Kennicutt--Schmidt relation shown in terms of the dependence of molecular gas depletion time $\tauH2 = \SH2/\SSFR$ on molecular gas surface density $\SH2$, with $\SSFR$ and $\SH2$ averaged on $1\kpc$ scale using a 2D Gaussian filter. The lines show the median relation, stacked over 11 snapshots between 500 and 600 Myr for $\fgas \sim 20\%$ and between 700 and 800 Myr for $\fgas \sim 40\%$, respectively. The shaded region shows the patch-to-patch variation for the run with suppressed CR diffusivity (16th--84th inter-percentile range). The scatter is similar in two other runs and therefore is not shown in the figure. The inner $R<1\kpc$ is excluded from the analysis, and molecular masses include the correction due to helium, assuming a helium mass fraction of 24\%. CR diffusivity suppression leads to a much flatter trend of $\tauH2$ at high $\SH2$, which is closer to the observed near-constant $\tauH2$. }
\end{figure}

The strong sensitivity of dense gas and SFR distributions to the CR diffusivity model implies that these distributions and the spatial correlations between dense gas and SFR can potentially be used to constrain CR modeling. One example of such a correlation is the observed near-linear relation between the SFR and molecular gas surface densities on $\sim$1 kpc scale in nearby star-forming galaxies, the so-called molecular Kennicutt--Schmidt relation \citep[KSR; e.g.,][]{Wong.Blitz.2002,Bigiel.etal.2008,Leroy.etal.2013,Bolatto.etal.2017}. The linearity of this relation implies that the depletion time of molecular gas is near-constant, $\tauH2 = \SH2/\SSFR \sim 2 \pm 1 \Gyr$, independent of the molecular gas surface density, $\SH2$. 

To explore the molecular KSR in our simulations, we select molecular gas using Equation (6) from \citet{Gnedin.Draine.2014} and adjust the strength of the UV radiation field, $U_{\rm MW}$, to produce a realistic total molecular gas masses.
$U_{\rm MW}=40$ results in total molecular masses between $\sim(1\text{--}2)\times10^9 \Msun$ in our simulations with $\fgas\sim20\%$ and therefore we adopt this value in all our runs.\footnote{The relevant value of the $U_{\rm MW}$ is set by the strength of UV field near the dense regions which is dominated by the local sources and can be significantly higher than the solar neighborhood value of $U_{\rm MW}=1$.} Although the normalization of $\SH2$ and $\tauH2$ is sensitive to the choice of $U_{\rm MW}$, we still can compare their relative trends and the differences between the runs. Also, the value of $U_{\rm MW}$ is expected to be higher in the high-$\fgas$ galaxy due to higher SFRs; however, we ignore this difference and use the same value of $U_{\rm MW}=40$ because we are only interested in comparing the trends of $\tauH2$ for different treatments of CR feedback. We also checked that the effect of CR feedback on the KSR slope described below remains qualitatively similar when we decrease or increase $U_{\rm MW}$ by a factor of 10, even though the slope itself does change.

The relation between $\tauH2$ and $\SH2$ in our simulations with different gas fractions is shown in Figure~\ref{fig:tauH2}.
As the top panel shows, in the $\fgas\sim20\%$ run, the effect of CRs on the slope of molecular KSR is rather weak: $\tauH2$ has a slight negative trend $\tauH2 \propto \SH2^{\beta}$, with $\beta \lesssim 0.1$ in all three runs. The effect on the slope becomes much stronger for the $\fgas\sim40\%$ galaxy, especially at high $\SH2$. While $\tauH2$ maintains its weak trend in the simulation with locally suppressed CR diffusivity, in the two other runs, it obtains a strong negative trend, $\tauH2 \propto \SH2^{-0.3}$, implying a noticeably superlinear molecular KSR, $\SSFR \propto \SH2^{1.3}$.

The linearity of molecular KSR and constant $\tauH2$ can be explained as a cancellation of the trends of the gas residence time in molecular regions and star formation efficiency integrated over this time interval \citep[see][for a detailed discussion]{Semenov.etal.2019}. In the simulations presented in \citet{Semenov.etal.2019}, this cancellation was a consequence of the efficient feedback with SN momentum boosted by a factor of 5 and the definition of the star-forming gas based on the turbulent virial parameter. As our results indicate, this physical picture remains qualitatively similar when we remove the SN momentum boosting and let CRs with suppressed diffusivity mediate stellar feedback. In contrast, in the runs without CRs and with constant CR diffusivity, feedback cannot counteract dense clump formation, and as a result, the trends of molecular gas lifetime and integrated star formation efficiency do not cancel any more and molecular KSR becomes superlinear.

\subsection{Gamma-ray luminosity} 
\label{sec:results:SFR-Lgamma}

\begin{figure}
\includegraphics[width=\columnwidth]{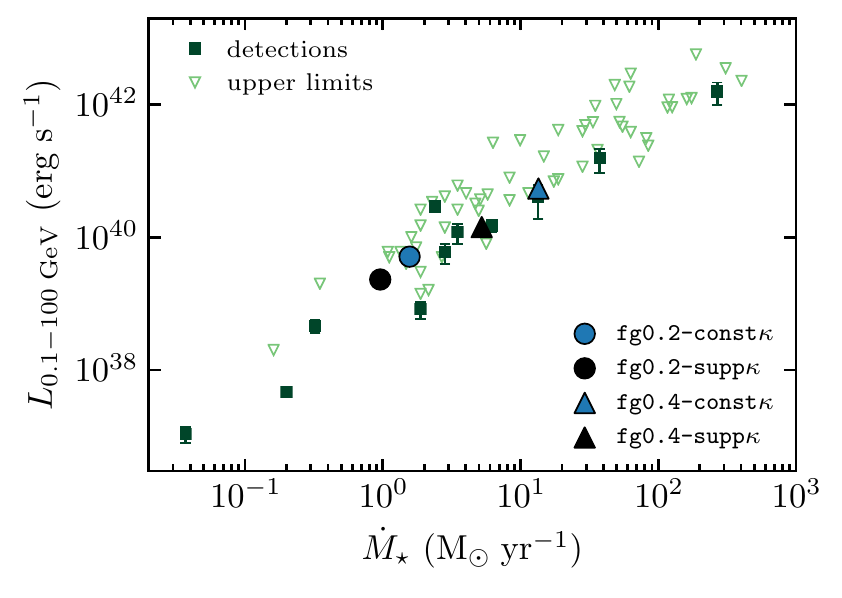}
\caption{\label{fig:Lgamma} Relation between the total $\gamma$-ray luminosity and SFR. Green squares and triangles show Fermi-LAT detections and upper limits, respectively \citep[the data are compiled from][]{Ackermann.etal.2012,Hayashida.etal.2013,Tang.etal.2014,Griffin.etal.2016,Peng.etal.2016,RojasBravo.Araya.2016}. The SFR and $L_\gamma$ are shown at the same times as the maps in Figures~\ref{fig:faceon-fg0.2} and \ref{fig:faceon-fg0.4}: $t=600\Myr$ for $\fgas = 20\%$ and $t=800\Myr$ for $\fgas = 40\%$. For consistency with the previous plots, the central $1\kpc$ is removed from the analysis. Adding the center results in an increase of both SFR and $L_\gamma$ by a factor of $\sim$1.5--2.}
\end{figure}

One of the existing constraints on the CR propagation models is the observed correlation between $\gamma$-ray luminosities, $L_\gamma$, and total SFRs. A major contribution to the observed $L_\gamma$ is the decay of $\pi^0$ produced in spallation reactions of CRs with the ISM. In galaxy simulations, accurate modeling of $L_\gamma$ is challenging because it strongly depends on the assumptions about the local CR energy spectrum and the structure of thermal gas and CR energy densities on unresolved scales. In our simulations, the latter is parameterized by the CR cooling suppression factor, $\floss$, that accounts for the fact that when CR diffusion is inhibited in star-forming regions, CRs spend most of the time in unresolved low-density regions within SN-blown bubbles (see Section~\ref{sec:sims:cr-losses}). Here we demonstrate that with our choice of CR diffusivity and loss suppression, our simulations agree with the observed SFR--$L_\gamma$ constraints.

To obtain the $\gamma$-ray luminosity, we assume that it is dominated by $\pi^0$ decay and compute the $\gamma$-ray emissivity in each cell as
\begin{equation}
\label{eq:Lgamma}
    \Lambda_\gamma = 5.6 \times 10^{-17}\left(\frac{\ecr}{\erg\cc}\right)\left(\frac{\neff}{\cc}\right)\ \ergs\cc,
\end{equation}
where $\ecr$ and $\neff = \floss n$ are the CR energy density and the effective density for CR losses in the cell, and the constant factor in front of this expression results from the integration of Equation (6) in \citet{Pfrommer.etal.2017b} between 0.1 and 100 GeV assuming the momentum spectral index of 2.05 and the low-momentum cutoff of $0.5\,m_{\rm p} c$. The total $\gamma$-ray luminosity is then obtained by integrating Equation~(\ref{eq:Lgamma}) over all cells within the galactocentric radii of $r=1\text{--}12\kpc$ and $|z|<2.5\kpc$ above and below the midplane: $\Lgamma = \int \Lambda_\gamma dV$. The central $1\kpc$ was removed for consistency with the previous plots; adding the center results in an increase of both SFR and $L_\gamma$ by a factor of $\sim$1.5--2 and thus does not change the conclusions.

The relation between $\gamma$-ray luminosity and SFR is shown in Figure~\ref{fig:Lgamma}. As the figure shows, all our simulations are consistent with the observed $L_\gamma$ for both constant diffusivity and $\kappacr$ suppressed near the injection sites. In the latter case, the loss suppression factor, $\floss$, is particularly important for producing realistic $L_\gamma$. The local suppression of CR diffusivity leads to CR accumulation in regions with high average densities, and if the unresolved density structures were not accounted for, the $\gamma$-ray fluxes would be larger by a factor of $\sim 5-10$ (see Appendix~\ref{app:var-factors} and Figure~\ref{fig:Lgamma-floss}). Although the model fluxes would in this case be higher than the formal observational estimates, the  uncertainties in $\kappacr$ and $\dot{M_\star}$ for observed galaxies are significant, and thus, it is not clear if such larger fluxes are inconsistent with observations.

\section{Discussion}
\label{sec:discussion}

The stability of gaseous disks is one of the key factors in galaxy evolution, especially in the early universe when galaxies were more compact, gas rich, and therefore more susceptible to gravitational instability. 
Indeed, a significant fraction of the observed $z>1$ galaxies and their local analogs exhibit UV-bright clumps of young stars \citep[e.g.,][]{Elmegreen.etal.2007,ForsterSchreiber.etal.2011,Guo.etal.2015,Guo.etal.2018,Livermore.etal.2015,Shibuya.etal.2016,Fisher.etal.2017}, which must be associated with dense gaseous clumps. 

In galaxy simulations, disk fragmentation and clump formation are strongly sensitive to the implementation of stellar feedback. 
Some cosmological simulations show violent fragmentation of unstable high-redshift disks that causes subsequent morphological transformations, which were proposed as a channel for galaxy quenching \citep[e.g.,][]{Dekel.etal.2009,Ceverino.etal.2010,Zolotov.etal.2015,Mandelker.etal.2017}. Other simulations, however, find that although massive clumps do form in high-redshift disks, these clumps are short lived and quickly dispersed by stellar feedback \citep[e.g.,][]{Genel.etal.2012,Buck.etal.2017,Oklopcic.etal.2017,Meng.Gnedin.2020}. 

Our results demonstrate that the local suppression of CR diffusion near the injection sites can prevent the fragmentation of the gaseous disk even when the disk is globally unstable. This effect can be an important channel for the regulation of star formation that is complementary to the effect of CR-driven winds extensively discussed in the literature (see references in the Introduction). 

Indeed, in the CR simulations with local diffusion suppression, clump formation is prevented by the local pressure gradients associated with the recently injected CRs that can accumulate near the injection sites. At the same time, after local SN explosions stop, CRs can escape and quickly build an extended vertical pressure gradient that can accelerate galactic winds.

Interestingly, the effect of CRs on fragmentation is expected to be more important for more massive and gas-rich galaxies, for which wind driving becomes less efficient. Thus, CRs can regulate galaxy formation via two complementary mechanisms: by driving galactic winds at low galaxy masses and suppressing disk fragmentation and formation of dense gas at high masses.

The effect of CR feedback on disk stability was previously pointed out in other numerical studies \citep[e.g.,][]{Salem.Bryan.2014,Pfrommer.etal.2017}. These studies showed that CRs can significantly increase the midplane ISM pressure and thicken the gaseous disk, which leads to disk stabilization and suppression of dense gas and star formation. Note, however, that just like the effect on wind driving, the effect on the midplane pressure is also complementary to the suppression of clump formation. Indeed, as our results demonstrate, CR feedback with constant diffusivity does improve global disk stability; however, it cannot prevent runaway fragmentation of unstable disks. In contrast, when the CR diffusion coefficient is reduced near the injection sites, disk fragmentation is suppressed due to the small-scale CR pressure gradients that counteract clump formation locally.

The effect of CR diffusion suppression on clump formation is qualitatively similar to that of strong stellar feedback: unstable disks can form clumps but these clumps are quickly dispersed by a combined effect of local CR pressure gradients and momentum injection by SNe. The resulting suppression of disk fragmentation can help to stabilize gaseous disks at high redshifts and thus to explain the existence of massive, dynamically cold disks that can form by $z \gtrsim 4$ according to recent discoveries \citep{Neeleman.2020,Rizzo.2020}. The abundance and lifetimes of massive clumps in such high-redshift disks are strongly sensitive to the degree of CR diffusion suppression. This sensitivity can be potentially exploited to constrain CR propagation models by comparing predictions of cosmological simulations with CR feedback to the abundance of UV-bright clumps in observed galaxies.

The effect of CRs on dense gas formation can also be important in lower-mass galaxies that have more stable gas disks. In such galaxies, local CR pressure gradients may affect the clustering of dense gas and young stars and the mass functions of giant molecular clouds and star clusters, which were shown to be sensitive probes of star formation and feedback modeling \citep[e.g.,][]{Li.etal.2018,Li.etal.2020,Semenov.etal.2018,Buck.etal.2019}. Thus, such small-scale statistics can also potentially constrain CR propagation in galaxy disks.
In addition, a qualitatively similar effect of CRs on the dense gas structure was also demonstrated on scales of individual star-forming regions by \citet{Commercon.etal.2019}, who showed that CRs with a small diffusion coefficient can inhibit the development of thermal instability in simulations of multiphase ISM turbulence.

Finally, it is worth noting that the effect of CR feedback with locally suppressed diffusion is qualitatively similar to the ``delayed cooling’’ or ``blastwave’’ feedback prescriptions used in some galaxy formation simulations \citep[e.g.,][]{Thacker.Couchman.2000,Stinson.etal.2006,Governato.etal.2007,Agertz.etal.2011}. In such prescription, gas cooling is delayed for a certain period of time after SN energy is injected in the form of thermal energy. This leads to a build-up of strong local pressure gradients that can disperse dense regions. 

However, theoretical models of SN-driven bubbles show that most of the thermal energy is in fact radiated away on timescales much shorter than the commonly assumed duration of suppression, and thus, theoretical basis for the delayed cooling models was not clear. If the microphysics of CR propagation and interaction with surrounding plasma does indeed lead to diffusion and cooling suppression  in star-forming regions, this can provide a physical basis for such models. 

There are also interesting differences between CR diffusion and cooling suppression and the standard delayed cooling of thermal energy. First, only a fraction of the SN energy can be converted to CRs and be contained near the SN bubbles. Second, after these bubbles are disrupted, CRs do not radiate away but escape into the ISM and inner halo, where they can provide additional pressure support to the disk or facilitate the acceleration of galactic wind. Thus, the effects of such a CR propagation model on galaxy evolution can be qualitatively different than in the simulations with the delayed cooling feedback and are worth exploring in the future.

\section{Summary and conclusions}
\label{sec:summary}

Observations of the $\gamma$-ray emission around young star clusters and isolated SN remnants suggest that the CR diffusion coefficient near their acceleration sites can be suppressed by a large factor, up to several orders of magnitude (see Section~\ref{sec:crsupp:obs}). Such suppression is also supported by analytical and numerical studies that show that CRs escaping from the acceleration sites can be self-confined in the extended regions around the shocks as a result of driving resonant and nonresonant modes via the streaming instability (see Section~\ref{sec:crsupp:theory}).

In this study, we explored the effects of CR diffusion suppression in star-forming regions on galaxy evolution by using simulations of isolated disk galaxies with different gas mass fractions: $\fgas\sim20\%$, which represents a typical \Lstar~galaxy at $z=0$ and $\fgas\sim40\%$, a gas-rich gravitationally unstable galaxy more typical for earlier stages of galaxy evolution. To isolate the effects of local CR diffusion suppression from other effects of CR feedback, we resimulated both galaxies in three regimes: (i) no CR feedback at all, (ii) CR feedback with constant and isotropic diffusion with the coefficient of $\kappacr=10^{28}\;\cm2s$, and (iii) CR feedback with $\kappacr$ suppressed in the regions where SN feedback is ongoing and where CRs are injected.

Our main results can be summarized as follows:
\begin{enumerate}
    \item CR feedback in the model with $\kappacr=\rm const$ can marginally improve the global disk stability by increasing the midplane pressure of the disk. As was found in previous studies, it enhances the overall effects of feedback for a given SFR.  
    However, such feedback cannot prevent the formation of dense star-forming clumps when the gas disk is gravitationally unstable  because CRs quickly diffuse away from dense regions of the ISM. 

    \item Local suppression of CR diffusion in star-forming regions, on the other hand, can efficiently suppress the formation of dense clumps. The accumulation of CRs and the build-up of their pressure due to the suppression of their propagation in these regions create large local pressure gradients that prevent clump formation, even when the disk is violently unstable globally.

    \item The suppression of clump formation in the model with locally reduced CR diffusion also leads to a decrease of the SFR. The magnitude of the SFR suppression is  similar to that due to the effect of strong stellar feedback that is often achieved in galaxy simulations via increasing local star formation efficiency or energy and momentum injection per SN, but achieved with CRs with small efficiency and without boosting SN energy or momentum.
    
    \item Interestingly, a less clumpy distribution of dense gas and SFR leads to a near-linear relation between molecular gas and SFR surface densities on kiloparsec scales even for a gas-rich, highly unstable disk, while the models with constant CR diffusivity or no CRs at all result in a superlinear relation.

    \item The suppression of CR diffusion in star-forming regions does not significantly alter the average midplane pressure profiles and properties of the gas and outflows in the inner halo in comparison with the constant $\kappacr$ model. 

    \item All our CR feedback models are consistent with the observed correlation between SFR and $\gamma$-ray luminosity, $L_\gamma$. To achieve such an agreement, the computation of $L_\gamma$ must account for the fact that in the regions with suppressed diffusivity, CRs predominantly occupy multiphase diffuse superbubbles, leading to a significant suppression of CR losses and $\gamma$-ray production.
\end{enumerate}

Our results demonstrate that the local suppression of CR transport near the injection sites can have large, qualitative effects on the morphology of star-forming galaxies, especially for gas-rich unstable disks. The magnitude of this effect is of course sensitive to the parameters of the model: the degree of diffusion suppression and the complex structure of gas and CRs on unresolved scales that determines the rates of CR losses. 

The sensitivity of disk morphology and stability to these parameters implies that such CR models can be constrained by the observations of the clumps in high-redshift galaxies, which motivates the further exploration of such models in cosmological simulations. At the same time, it would 
be extremely interesting to model the effects of CR suppression on the structure of the interstellar medium in high-resolution simulations of the ISM patches and individual star-forming regions with CR acceleration and more detailed models of propagation suppression near the shocks.

\acknowledgements

We would like to thank Mateusz Ruszkowski and members of the galaxy formation group at UChicago for many useful discussions. 
We also thank the anonymous referee for the constructive and thoughtful review that helped to improve this paper.
This work was supported by the NSF grant AST-1714658.
Support for V.S. was also provided by NASA through the NASA Hubble Fellowship grant HST-HF2-51445.001-A awarded by the Space Telescope Science Institute, which is operated by the Association of Universities for Research in Astronomy, Inc., for NASA, under contract NAS5-26555.
A.K. was also supported by the NSF grant AST-1911111. 
D.C. was also partially supported by NASA (grants NNX17AG30G, 80NSSC18K1726, and 80NSSC20K1273) and by NSF (grants  AST-1909778 and PHY-2010240).
The simulations presented in this paper have been carried out using the Midway cluster at the University of Chicago Research Computing Center, which we acknowledge for support. Analyses presented in this paper were greatly aided by the following free software packages: {\tt yt} \citep{yt}, {\tt NumPy} \citep{numpy_ndarray}, {\tt SciPy} \citep{scipy}, {\tt Matplotlib} \citep{matplotlib}, and \href{https://github.com/}{GitHub}. We have also used the Astrophysics Data Service (\href{http://adsabs.harvard.edu/abstract_service.html}{ADS}) and \href{https://arxiv.org}{arXiv} preprint repository extensively during this project and writing of the paper.

\appendix

\section{Entropy-Conserving Scheme for Cosmic Ray Modeling}
\label{app:entropy}

\begin{figure*}
\includegraphics[width=\textwidth]{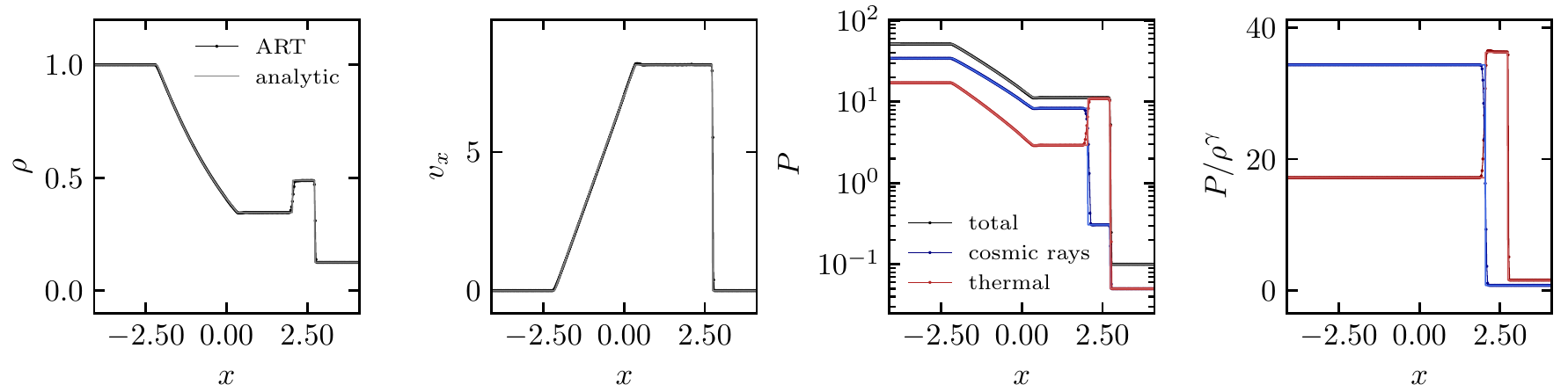}
\caption{\label{fig:shocktube} The shock-tube test with CRs using the initial conditions from \citet{Pfrommer.etal.2017}: $(\rho,v_x,\Pth,\Pcr) = (1,0,17.172,34.344)$ and $(0.125,0,0.05,0.05)$ for the left and right initial states, respectively. The results are plotted at $t=0.25$, and it agrees with the analytic solution from \citet{Pfrommer.etal.2017} shown with the thin lines. As the last panel shows, the entropy-based method ensures CR entropy conservation across the shock, and all energy dissipated by the shock is correctly converted into thermal energy.}
\end{figure*}

In this section, we describe how we solve the most basic part of the CR evolution, Equation (\ref{eq:ecr}), advection, and the $PdV$ work:
\begin{equation}
\label{eq:adv-pdv}
    \frac{\partial \ecr}{\partial t} + \nabla (u \ecr) = -\Pcr \nabla u,
\end{equation}
with $\Pcr = (\gcr - 1) \ecr$ and $\gcr=4/3$.
Although we will focus on modeling CRs, we use the same method to follow other nonthermal energies, in particular, unresolved turbulent energy (see Section~\ref{sec:sims:overview}). 

An equation similar to Equation (\ref{eq:adv-pdv}) is in fact solved in many finite-volume galaxy formation codes to follow thermal energy as an independent fluid variable. While thermal energy can be computed as the difference between total and kinetic energy, $\eth = \etot-\ekin = \etot-\mathbf{p}^2/(2\rho)$, the necessity to model $\eth$ separately from $\etot$ arises in highly supersonic flows, when $\eth$ can become comparable to or smaller than the truncation error of $\etot$ and thus become highly inaccurate  \citep{Ryu.etal.1993,Bryan.etal.1995}. Having two independent ways to estimate $\eth$, one also needs to define the criteria whether the independently followed $\eth$ or $\etot-\ekin$ should be used in a given cell to compute pressure, temperature, cooling rate, etc.

The advection and $PdV$ work of CRs and other nonthermal energies can (and should) be modeled using the same method as that for the thermal energy. One important modification that should be made is how $\eth$ and nonthermal energies are synchronized with $\etot$: the difference $\etot-\ekin$ now corresponds to the sum of thermal and all nonthermal energies, so one needs to decide how to partition this difference. At shocks, this difference contains the adiabatic change of thermal and nonthermal energies and the energy dissipated by the shock, and therefore the choice of partitioning will depend on the expected behavior of nonthermal energies across shocks. Real shocks can generate CRs and turbulence, which can be taken into account in the partitioning scheme. However, in the absence of a subgrid model for such generation, the conservative choice is to assume that all energy dissipated by shocks is thermalized,
\begin{equation}
\label{eq:eth-sync}
    \eth = \etot-\frac{\mathbf{p}^2}{2\rho} - \ecr,
\end{equation}
while the nonthermal energies, in this case $\ecr$, change adiabatically.

The original implementation of thermal and nonthermal energies in the ART code was based on the method proposed by \citet{Bryan.etal.1995}, where Equation~(\ref{eq:adv-pdv}) is solved directly by advecting $\ecr/\rho$ as a passive scalar and adding the $PdV$ work as a source term. In the ART code, $\eth$ was then synchronized with $\etot$ in the regions where $\eth/(\etot-\ekin-\ecr)>10^{-3}$ using Equation~(\ref{eq:eth-sync}). While this method performs well for modeling thermal energy only, we find that it does not ensure the adiabatic change of nonthermal energies across shocks. In the shocked regions, the $PdV$ source-term consists of both the adiabatic part and the energy dissipated by the shock, which are difficult to disentangle. As a result, using this method to advance $\ecr$ results in the generation of nonthermal entropy at shocks. 

To enforce nonthermal entropy conservation across shocks, we switched to the method proposed by \citet{Ryu.etal.1993}, where the advection and $PdV$ work are modeled by solving a conservative equation for modified entropy, $\rho\Scr = \Pcr/\rho^{\gcr-1}$:
\begin{equation}
\label{eq:entropy}
    \frac{\partial \rho\Scr}{\partial t} + \nabla (u \rho\Scr) = 0.
\end{equation}
This expression can be derived by combining Equation~(\ref{eq:adv-pdv}) with the continuity equation. The quantity $\Scr = \Pcr/\rho^\gcr$ is a monotonic function of gas entropy per unit mass, and thus, this method ensures entropy conservation by passing $\Scr$ from cell to cell as a passive scalar.

To ensure consistency between thermal and nonthermal energies, we use the same entropy-based method to advance thermal energy. However, this way of modeling $\eth$ is valid only outside shocked regions because shocks do generate thermal entropy. To capture this generation, $\eth$ must be reset from $\etot$ in the shocked regions. To identify such regions, we follow \citet{Springel.2010} and synchronize $\eth$ and $\etot$ in the cells where the largest Mach number of the shocks present in the Riemann solutions on its interfaces exceeds a threshold $M_{\rm crit} = 1.1$. Although the relation between the shocks in the Riemann solutions and the real shocks is nontrivial, the total entropy generated by the real shock accumulates from the increments produced by the ``Riemann shocks'' on the interfaces resolving the real shock, and these increments become significant at Mach numbers $\gtrsim$1.1 \citep[see Figure 12 in][]{Springel.2010}.

As was also pointed out by \citet{Springel.2010}, enforcing entropy conservation forfeits the energy conservation of the scheme. In real applications, however, the total energy is not conserved anyway due to, e.g., cooling/heating and star formation feedback processes. At the same time, we find that in idealized tests, the entropy-based scheme performs either comparably or better than the energy-based scheme.

One of the tests of our entropy-based method is shown in Figure~\ref{fig:shocktube}, which compares a shock-tube problem with CRs with the analytic solution from \citet{Pfrommer.etal.2017}. The last panel, in particular, shows the entropy of thermal gas (red) and CRs (blue). CR entropy is conserved across the shock and changes only at the contact discontinuity, while the energy dissipated by the shock is correctly converted to thermal energy in agreement with the analytic solution.

As a side note, the entropy-based modeling of thermal and nonthermal energy also provides an approximate but very cheap way to implement the generation of nonthermal energies by shocks, such as CR acceleration, without requiring explicit shock finding. Indeed, as pointed out above, such generation can be implemented by appropriately partitioning $\etot-\ekin$ between $\eth$ and nonthermal energies in the shocked regions instead of using Equation~(\ref{eq:eth-sync}). After each hydro step, $e_{\rm diss} \equiv \etot-\ekin-\eth-\ecr$ corresponds to the total energy dissipated by shocks in each cell during the step. Therefore, to convert a fraction $\zeta$ of the dissipated energy into CRs, one just needs to add $\zeta\times e_{\rm diss}$ to $\ecr$ and $(1-\zeta)\times e_{\rm diss}$ to $\eth$. More details and additional tests will be provided in a forthcoming paper.

\section{Diffusion solver test}
\label{app:diffusion}

\begin{figure}
\includegraphics[width=\columnwidth]{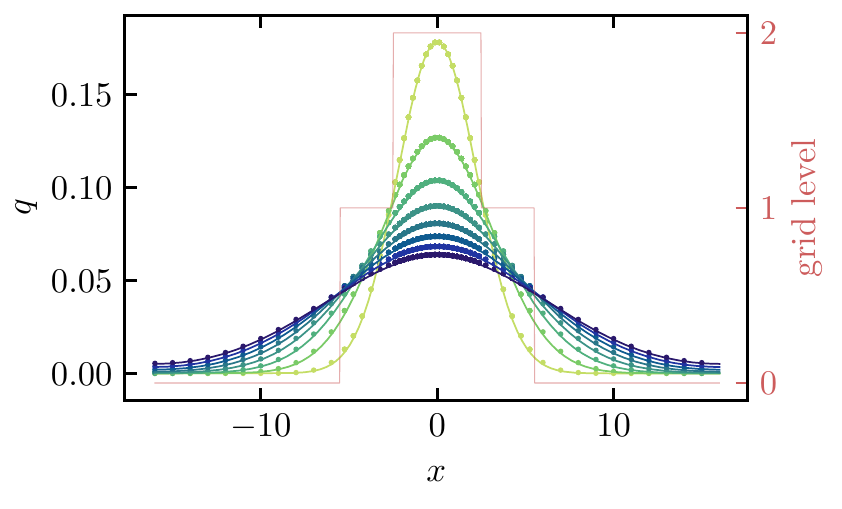}
\caption{\label{fig:diffusion} Comparison of the 1D diffusion test (points) with the analytical solution (lines). Colors from green to blue show the outputs at 5, 10, 15, ..., $40\;t_{\rm diff,cell}$, where $t_{\rm diff,cell} = \Delta_0^2/(2 \kappa)$ is the cell diffusion time at the lowest refinement level. The thin red line indicates the grid refinement levels. The cells on the highest level make two diffusion subcycles per step (as also typically the case for our galaxy runs), while the other two levels are advanced without diffusion subcycling. Neither refinement nor subcycling introduces any strong artifacts in the solution. }
\end{figure}

CR and turbulent diffusion terms are solved using an explicit Forward Time Centered Space scheme \citep[e.g.,][]{numerical-recipes}. While this scheme puts a stringent constraint on the time step, $\Delta t \propto \Delta x^2$, for our rather moderate resolution of $\Delta x=40\pc$, it is not prohibitive. Nevertheless, we perform subcycling of the diffusion solver over the hydrodynamic step to speed up the computation. In the galaxy simulations presented in this paper, the maximum number of required subcycles was two.

Figure~\ref{fig:diffusion} shows a one-dimensional point-source diffusion test with mesh refinement and subcycling.
In this test, the CR energy is initialized in a single cell, and gas density is set to an arbitrary large value ($\rho = 10^{30}$) so that the advection terms become negligible and the evolution of CRs is fully diffusive. The figure compares the evolution of CR energy density normalized by the total initial CR energy with the analytic solution that accounts for the first periodic images of the source:
\begin{align*}
    q(x,t) &= q_1(x-L_{\rm box},t)+q_1(x,t)+q_1(x+L_{\rm box},t),\\
    q_1(x,t) &= \frac{1}{\sqrt{4 \pi \kappa t}} e^{-x^2/(4 \kappa t)}.
\end{align*}
As the figure shows, neither subcycling nor refinement boundaries introduce noticeable artifacts.

\section{Variation of CR diffusion suppression and effective density for CR losses}
\label{app:var-factors}

\begin{figure}
\includegraphics[width=\columnwidth]{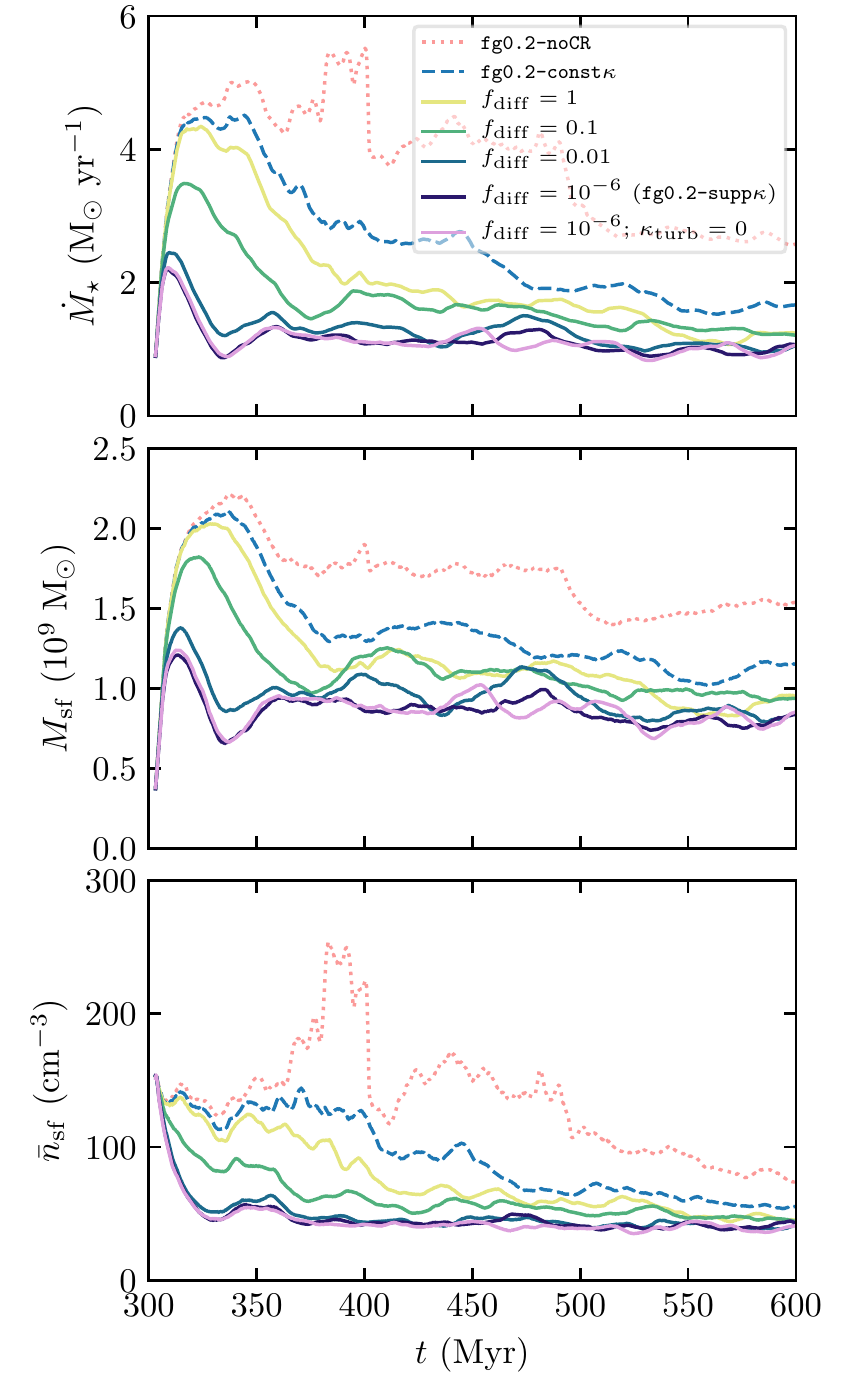}
\caption{\label{fig:var-diff} Effect of varying $\fdiff$ while keeping $\floss=0.01$ as in our fiducial run with suppressed CR diffusivity. The red dotted and blue dashed lines show the runs without CRs and with constant diffusivity (i.e., with $\fdiff=1$ and $\floss=1$). The violet line shows the simulation where we switched off turbulent diffusion ($c_\kappa = 0$ in Equation~\ref{eq:kappa-sgst}). The effect increases for stronger diffusion suppression (smaller $\fdiff$) and saturates at $\fdiff\lesssim0.01$. }
\end{figure}

\begin{figure}
\includegraphics[width=\columnwidth]{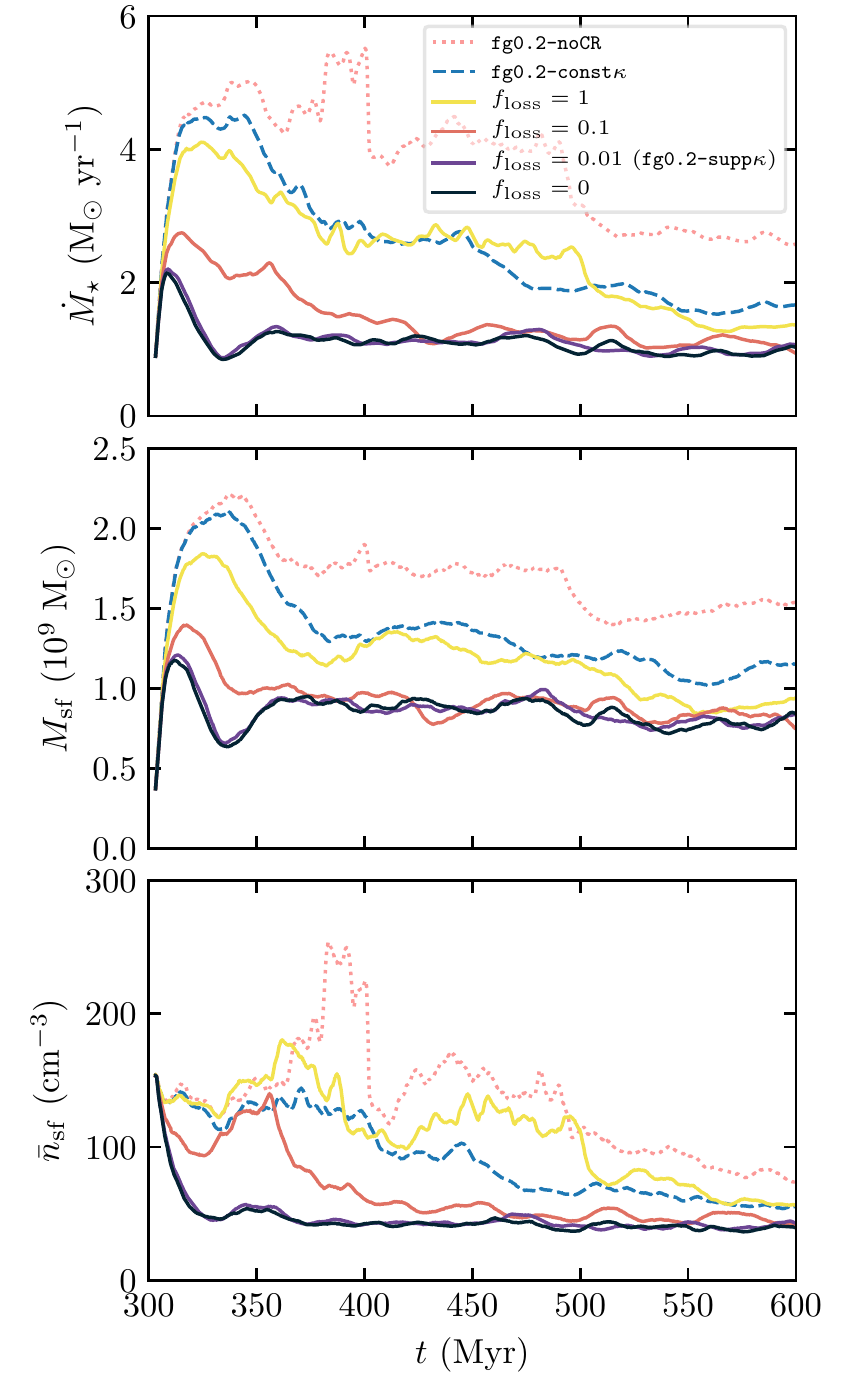}
\caption{\label{fig:var-loss} Effect of varying $\floss$ while keeping $\fdiff=10^{-6}$ as in our fiducial run with suppressed CR diffusivity. The red dotted and blue dashed lines show the runs without CRs and with constant diffusivity (i.e., with $\fdiff=1$ and $\floss=1$). The effect increases for stronger suppression of losses (smaller $\floss$) and saturates at $\floss\lesssim0.05$. }
\end{figure}

\begin{figure}
\includegraphics[width=\columnwidth]{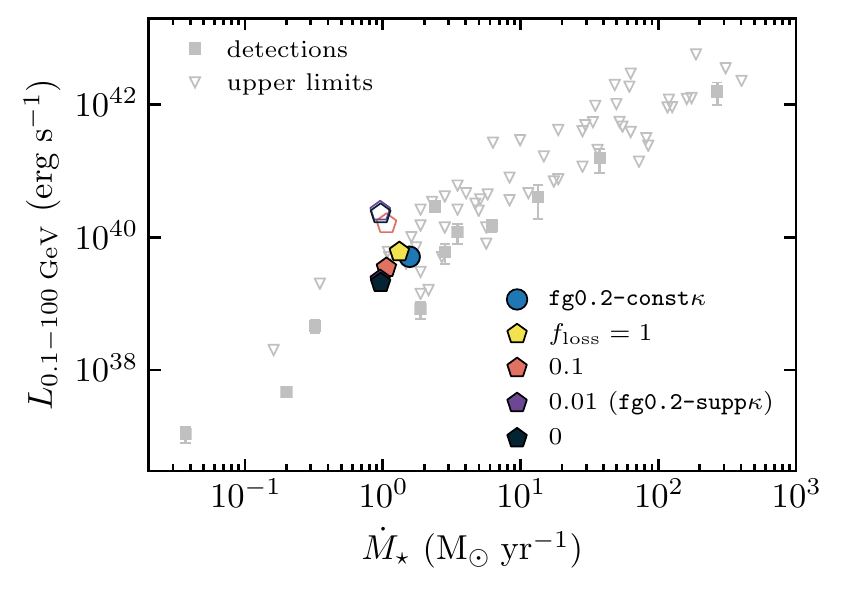}
\caption{\label{fig:Lgamma-floss} The effect of $\floss$ variation on the relation between SFR and $\gamma$-ray luminosity. The filled polygons show the $\Lgamma$ computed consistently with the CR losses with the color corresponding to the $\floss$ value, while the open polygons show the $\Lgamma$ computed without accounting for the $\floss$ factor, i.e., assuming $\floss=1$ in all cases. The gray squares and triangles show Fermi-LAT detections and upper limits, respectively (the references are provided in the caption of Figure~\ref{fig:Lgamma}).}
\end{figure}

\begin{figure*}
\centering
{\large Moderate gas fraction galaxy, $f_{\rm gas} = 20\%$:}\\
\includegraphics[width=\textwidth]{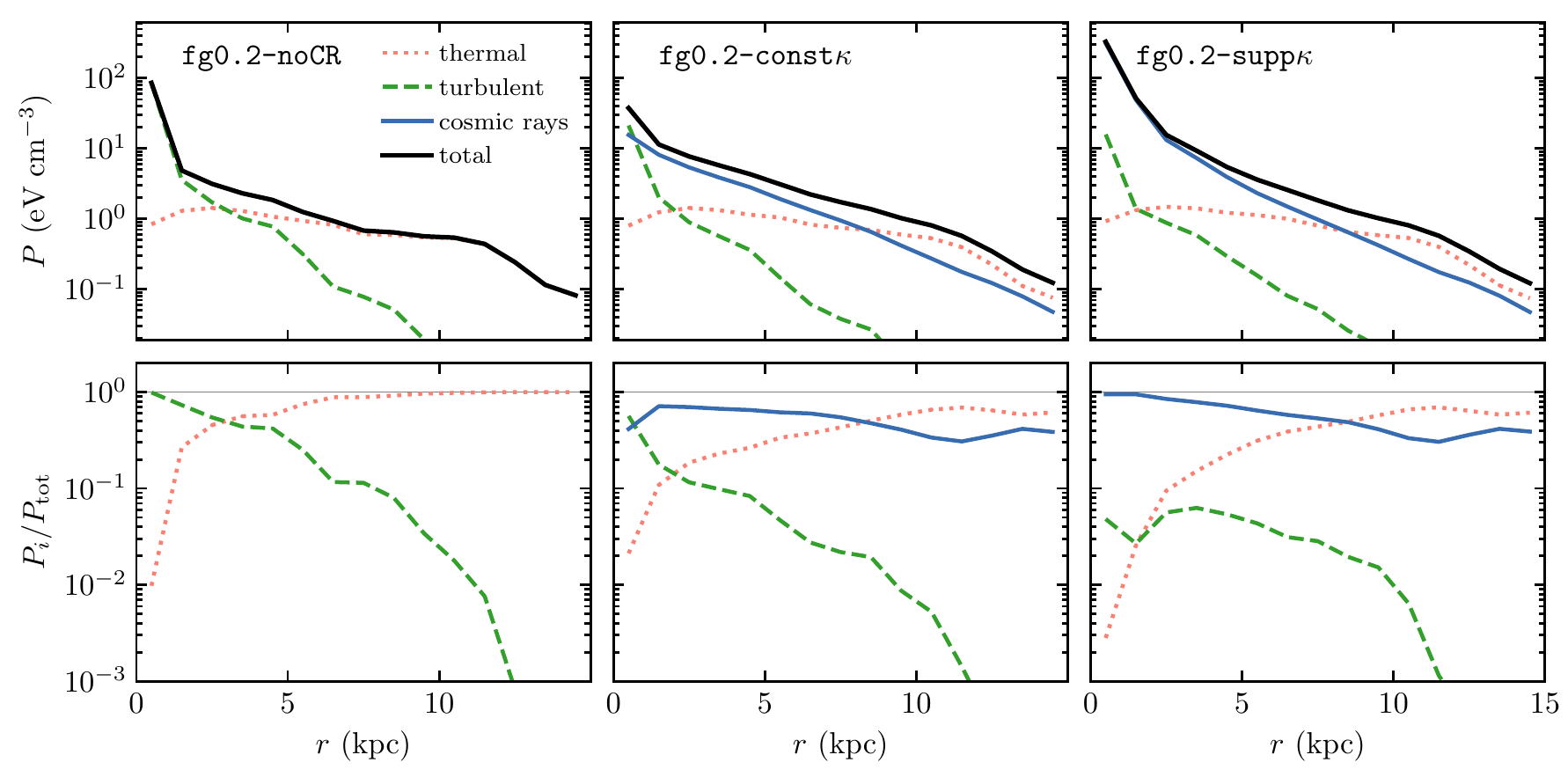}\\
{\large High gas fraction galaxy, $f_{\rm gas} = 40\%$:}\\
\includegraphics[width=\textwidth]{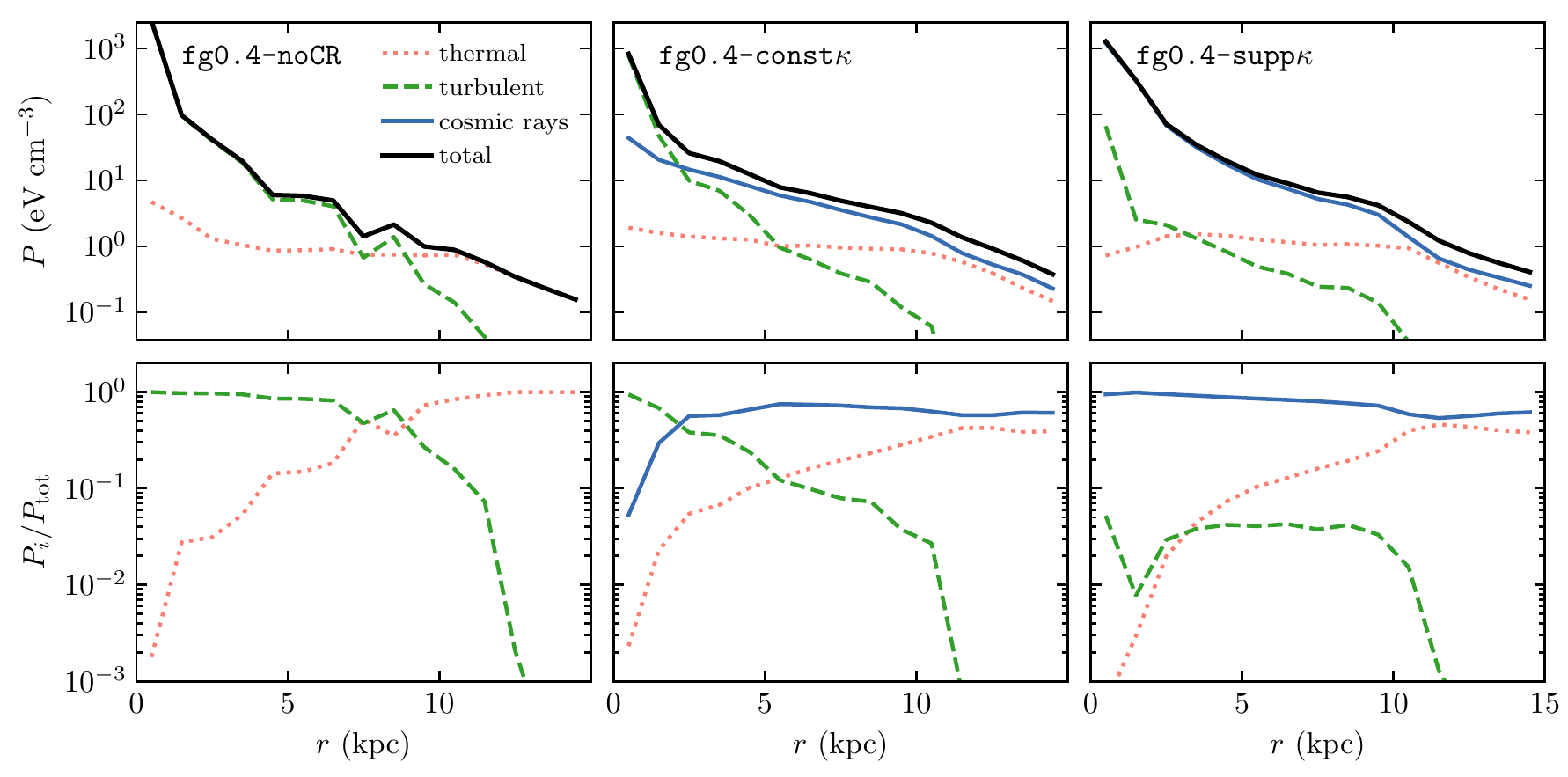}
\caption{\label{fig:pressure} Radial profiles of the midplane pressure weighted by area and fractions of thermal, turbulent, and CR pressures in the total midplane pressure. The top and bottom sets of panels show the results for galaxy simulations with the gas fraction of $\fgas \sim 20\%$ and 40\%, respectively. The profiles of thermal, turbulent, and CR pressure are stacked over 11 snapshots between 500 and 600 Myr for $\fgas \sim 20\%$ and between 700 and 800 Myr for $\fgas \sim 40\%$, respectively, with the lines showing the medians. The total pressure is computed as a sum of median profiles of all pressure components.}
\end{figure*}

In this section, we explore the effect of various diffusion suppression factors in the vicinity of SNe, $\fdiff = \kappa_{\rm cr,SN}/\kappa_{\rm cr,0}$, and the effective density for CR losses in the same regions parameterized as $\neff = \floss \; n_{\rm cell}$. In our simulations presented in the main part of the paper, these parameters are $\fdiff=1$, $\floss=1$ for the run with constant CR diffusivity (\texttt{fg0.2-const$\kappa$}) and $\fdiff=10^{-6}$, $\floss=0.01$ in the run with locally suppressed CR diffusivity (\texttt{fg0.2-supp$\kappa$}).

Figure~\ref{fig:var-diff} shows the effect of varying $\fdiff$ while keeping $\floss=0.01$ as in our fiducial run with suppressed CR diffusivity. Stronger suppression of diffusivity near the injection sites leads to a stronger effect on the SFR and the amount and densities of star-forming gas. The effect, however, saturates at $\fdiff<0.01$ because the time to diffuse away from the regions with suppressed $\kappacr$ becomes longer than the duration of suppression, $40\Myr$ (see Section~\ref{sec:sims:diffusion} and Equation~\ref{eq:diff-time}). As also detailed in Section~\ref{sec:sims:diffusion}, at small $\fdiff$ the CR diffusion is dominated by the unresolved turbulent advection. However, its effect on the results is negligible as we explicitly show in the figure by switching off turbulent diffusion ($c_{\kappa} = 0$ in Equation~\ref{eq:kappa-sgst}) in the run with $\fdiff=10^{-6}$.

The bottom two panels highlight the effect of varying $\fdiff$ on the amount and density of star-forming gas. As detailed in Section~\ref{sec:results:sfr}, the higher $\Msf$ and $\bar{n}_{\rm sf}$ in the runs without CR diffusion suppression are due to the formation of dense star-forming clumps, especially at early times, when the difference between the runs is the largest. As the figure demonstrates, both $\Msf$ and $\bar{n}_{\rm sf}$, and therefore the abundance of dense clumps, monotonically decrease with decreasing $\fdiff$ until the effect saturates at $\fdiff \sim 0.01$.

Figure~\ref{fig:var-loss} shows the effect of varying $\floss$ while keeping the fiducial $\fdiff=10^{-6}$. The effect becomes stronger at smaller $\floss$ and, similarly to the effect of diffusion suppression, it quickly saturates as the CR loss time become longer than the duration of suppression (see Section~\ref{sec:sims:cr-losses} and Equation~\ref{eq:loss-time}).
Figure~\ref{fig:Lgamma-floss} also shows the effect of $\floss$ variation on the SFR--$\Lgamma$ relation. The $\gamma$-ray luminosity is only weakly sensitive to the $\floss$ value as in all presented cases the galaxy remains close to the colorimetric limit. To compute $\Lgamma$ consistently with the CR losses adopted in the simulation, it is important to account for $\floss$ in the calculation of  $\Lgamma$. As the empty polygons show, not accounting for $\floss$ results in an order of magnitude higher $\Lgamma$ in runs with $\floss<1$. However, given the significant uncertainties in $\kappacr$ and $\dot{M_\star}$ for observed galaxies, it is not clear that even such high fluxes are inconsistent with observations.

Another notable conclusion from Figures~\ref{fig:var-diff} and \ref{fig:var-loss} is that the strong effect on the SFR and dense gas formation in our fiducial run with suppressed diffusivity results from the suppression of both CR diffusivity and losses. Indeed, as runs with $\fdiff=1$ or $\floss=1$ show, if only diffusivity or losses are suppressed but not both, the result is closer to the simulation with no suppression of diffusivity or losses at all (\texttt{fg0.2-const$\kappa$}). This is because CRs either quickly escape from dense gas when there is no diffusivity suppression ($\fdiff=1$) or quickly lose their energy due to too high loss rates in dense regions ($\floss=1$).

\section{Radial pressure profiles}
\label{app:pressure}

Figure~\ref{fig:pressure} shows the radial profiles of the midplane pressure for different CR feedback models for simulations with $\fgas \sim 20\%$ (top set of panels) and $40\%$ (bottom set of panels). 

In all $\fgas \sim 20\%$ runs, both thermal and turbulent pressures remain approximately the same, with thermal pressure dominating over turbulence at $R>2\kpc$. 
In the presence of CR and turbulent pressure support, the thermal pressure, $\Pth \propto nT$, is set by the net heating and cooling and is roughly equal in diffuse interarm gas ($n \sim 1\cc$ and $T \sim 10^4\K$) and in dense regions ($n \sim 100\cc$ and $T \sim 100\K$). As a result, the average midplane thermal pressure is almost independent of radius and does not change much between the runs with different $\fgas$: only the partitioning of gas between the warm and cold phases changes. On the other hand, CR pressure depends strongly on local SFR, and therefore, it increases toward the disk center following the roughly exponential radial profile of the SFR. Interestingly, for $\fgas\sim20\%$, thermal and CR pressures become equal around the solar radius, $R \sim 8 \kpc$, which is consistent with the observed equipartition at the $\sim 1 \eVcc$ level in the local ISM \citep[e.g.,][]{Grenier.etal.2015}. When $\fgas$ is increased from 20\% to 40\%, the SFR increases by a factor of $\sim 5$ (see Figures~\ref{fig:sfh-fg0.2} and \ref{fig:sfh-fg0.4}), and the CR pressure raises by a similar factor, becoming dominant throughout the disk.

In the run with suppressed CR diffusivity, the midplane pressure increases only in the very center of the disk, where the CRs ``trapped'' near the injection sites dominate (see the fourth panel in the bottom row of Figure~\ref{fig:faceon-fg0.2}). In most of the disk, the midplane pressure does not change significantly because it is dominated by the diffuse CR pressure component that is insensitive to the CR diffusivity suppression (see Section~\ref{sec:results:maps}). 

In the $\fgas \sim 40\%$ runs, the midplane pressure profiles are qualitatively similar to those with $\fgas \sim 20\%$, except that the contribution of turbulent and CR pressure becomes larger than that of the thermal pressure.

\bibliographystyle{aasjournal}
\bibliography{}

\end{document}

%% file: citation_fix.tex
\usepackage{etoolbox}
\makeatletter
\patchcmd{\NAT@citex}
  {\@citea\NAT@hyper@{\NAT@nmfmt{\NAT@nm}\NAT@date}}
  {\@citea\NAT@nmfmt{\NAT@nm}\NAT@hyper@{\NAT@date}}
  {}
  {}
\patchcmd{\NAT@citex}
  {\@citea\NAT@hyper@{%
     \NAT@nmfmt{\NAT@nm}%
     \hyper@natlinkbreak{\NAT@aysep\NAT@spacechar}{\@citeb\@extra@b@citeb}%
     \NAT@date}}
  {\@citea\NAT@nmfmt{\NAT@nm}%
   \NAT@aysep\NAT@spacechar%
   \NAT@hyper@{\NAT@date}}
  {}
  {}
\patchcmd{\NAT@citex}
  {\@citea\NAT@hyper@{%
     \NAT@nmfmt{\NAT@nm}%
     \hyper@natlinkbreak{\NAT@spacechar\NAT@@open\if*#1*\else#1\NAT@spacechar\fi}%
       {\@citeb\@extra@b@citeb}%
     \NAT@date}}
  {\@citea\NAT@nmfmt{\NAT@nm}%
   \NAT@spacechar\NAT@@open\if*#1*\else#1\NAT@spacechar\fi%
   \NAT@hyper@{\NAT@date}}
  {}
  {}
\makeatother

%% file: defs.tex
\def\Msf{{M}_{\rm sf}}
\def\Mg{{M}_{\rm g}}
\def\MH2{{M}_{\rm H_2}}

\def\SFR{\dot{M}_\star}

\def\rhoSFR{{\dot{\rho}}_\star}
\def\rhoH2{\rho_{\rm H_2}}
\def\epsff{\epsilon_{\rm ff}}

\def\avir{\alpha_{\rm vir}}

\def\cs{c_{\rm s}}
\def\st{\sigma_{\rm turb}}
\def\stot{\sigma_{\rm tot}}

\def\SH2{\Sigma_{\rm H_2}}

\def\SSFR{\dot\Sigma_\star}

\def\fH2{f_{\rm H_2}}
\def\fgas{f_{\rm g}}

\def\b{b}

\def\tff{t_{\rm ff}}

\def\tauH2{\tau_{\rm dep,H_2}}

\def\const{{\rm const}}
\def\Lstar{$L_\star$}

\def\Lambdacr{\Lambda_{\rm cr}}
\def\ecr{e_{\rm cr}}
\def\Pcr{P_{\rm cr}}
\def\gcr{\gamma_{\rm cr}}
\def\Scr{S_{\rm cr}}

\def\kappacr{\kappa_{\rm cr}}
\def\neff{n_{\rm eff}}
\def\Lgamma{L_\gamma}
\def\vA{v_{\rm A}}
\def\vst{v_{\rm st}}

\def\eturb{e_{\rm turb}}
\def\eth{e_{\rm th}}
\def\etot{e_{\rm tot}}
\def\ekin{e_{\rm kin}}

\def\Pth{P_{\rm th}}

\def\fdiff{f_{\rm diff}}
\def\floss{f_{\rm loss}}
\def\ncell{n_{\rm cell}}

\def\pc{{\rm\;pc}}
\def\kpc{{\rm\;kpc}}
\def\Myr{{\rm\;Myr}}
\def\Gyr{{\rm\;Gyr}}
\def\Msun{{\rm\;M_\odot}}

\def\Msunpc2{{\rm\;M_\odot\;pc^{-2}}}
\def\kms{{\rm\;km\;s^{-1}}}
\def\cc{{\rm\;cm^{-3}}}
\def\cm2s{{\rm\;cm^{2}\;s^{-1}}}
\def\ccs{{\rm\;cm^{3}\;s^{-1}}}
\def\ergs{{\rm\;erg\;s^{-1}}}
\def\erg{{\rm\;erg}}

\def\K{{\rm\;K}}
\def\eVcc{{\rm\;eV\;cm^{-3}}}